\documentclass[sigconf]{acmart}

\renewcommand\footnotetextcopyrightpermission[1]{} 
\usepackage{booktabs} 
\usepackage{amsmath,amssymb,amsfonts}
\usepackage{algorithm}
\usepackage{algorithmicx}
\usepackage{algpseudocode}
\usepackage{graphicx}
\usepackage{textcomp}
\usepackage{subfigure}
\usepackage{xcolor}
\usepackage{epstopdf}

\setcopyright{none}

\begin{document}
\title{Multi-hop Data Fragmentation in Unattended Wireless Sensor Networks}

\author{Keun-Woo Lim}
\affiliation{%
  \institution{Telecom Paristech}
  \city{Paris}
  \state{France}
  \postcode{75013}
}
\email{keunwoo.lim@telecom-paristech.fr}

\author{Katarzyna Kapusta}
\affiliation{%
  \institution{Telecom Paristech}
  \city{Paris}
  \state{France}
  \postcode{75013}
}
\email{katarzyna.kapusta@telecom-paristech.fr}

\author{Gerard Memmi}
\affiliation{%
  \institution{Telecom Paristech}
  \city{Paris}
  \state{France}
  \postcode{75013}
}
\email{gerard.memmi@telecom-paristech.fr}

\author{Woo-Sung Jung}
\affiliation{%
  \institution{Electronics and Telecommunications Research Institute}
  \city{Daejeon}
 \state{South Korea}
  \postcode{}
}
\email{woosung@etri.re.kr}


\begin{abstract}
In this work, we analyze the advantages of multi-hop data fragmentation in unattended wireless sensor networks (UWSN) and propose a lightweight protocol to achieve it. UWSN has recently become an important aspect in various areas of sensor networks where real-time data collection is difficult to manage. However, the characteristics of UWSN also poses new problems especially in data protection. For more efficient protection, data fragmentation has been proposed to fragment sensing data, which prevents attackers from successfully exploiting the data. However, there are currently minimal work on the strategies of the placement of fragments inside a sensor network. Through this work, we analyze the effects of multi-hop fragment dispersal in relation to effectiveness of data protection and energy consumption. Furthermore, we design a new routing algorithm suitable for the energy-efficient placement of data fragments in UWSN. We utilize simulation-based modeling and testbed implementation via FIT/IoT-Lab to prove the effectiveness of our work.
\end{abstract}

%
%


\maketitle

\section{Introduction}
In various applications that require usage of a wireless sensor network (WSN), it can be difficult to use the commonly assumed network environment of WSN, which is the interaction and multi-hop wireless communication between a sink node and multiple sensors. 
This is the case of hostile environments such as battlefields and inaccessible regions (e.g. fallout sites) where it could be difficult to collect sensing data and manage the network in real-time manner. 
In these specific cases, it is more preferable to utilize itinerant sinks to be able to collect sensing data per demand. %
In the near future, we can envision new Internet of Things (IoT) services, comprised of high-tech devices, such as automated unmanned aerial vehicles (UAVs) to act as sink nodes in a sensor network to be able to provide these functionalities. 
Such kind of sensor network architecture has been defined by researchers as the unattended wireless sensor networks (UWSN) \cite{Ma_2009}. 
%

In UWSN, one of the most important factors is the survivability of the data, which takes into account whether a data generated by a sensor node is successfully transferred to the itinerant sink node. 
This is especially critical in UWSN because the data must be stored in the unattended sensors before an itinerant sink arrives in the proximity to receive the data. 
During this time, the data can be a viable target to attackers, who are also mobile and either steal, modify, or erase the data from the sensor nodes \cite{Dipietro_2008}\cite{Dipietro_2009}. 
%

One of the potential technologies to enhance the survivability of the data is data fragmentation\cite{Cheng_2014}\cite{Kapusta_2017}\cite{Kapusta_2019}. 
The general purpose of data fragmentation is to fragment a data $k$ into $f_k$ fragments through a cryptographic means. 
An itinerant sink can decode the data by collecting $f_k^\prime$ number of fragments where $f_k^\prime \leq f_k$. 
If the $f_k^\prime$ is high enough, it becomes more difficult for an attacker to be able to successfully exploit the data, as it must attack multiple sensors to be able to collect $f_k^\prime$ fragments. 
Therefore, data fragmentation provides protection of the data itself, while also providing some flexibility even if some of the fragments are erased or not received by the itinerant sink node.
%

However, majority of the state-of-the-art in data fragmentation only consider the security aspects, as in how to fragment the data and defragment them again. 
They do not consider where to place the fragments in the sensor network, even though it is very likely that in a sensor network, there can be numerous number of deployed sensors, while the conditions of which sensors can receive a data fragment can be very dynamic. 
Moreover, for the sake of transferring data to remote sensors, multi-hop routing protocols are needed. 
The general architecture of UWSN does not need multi-hop routing, therefore adding a routing layer just for the sake of transferring data fragments induce considerable energy consumption and bandwidth consumption from control packet transmission. 
Even though some epidemic flooding methods are proposed to disseminate data to remote sensor nodes \cite{Dipietro_2011}\cite{Aliberti_2017}, these are methods used for replication of same data, which have different characteristics and objectives compared to data fragmentation. 
Therefore, it is difficult to use epidemic protocols for dissemination of data fragments without inducing heavy overhead, because fragmented data are different in detail.
%

In this paper, we look at some methods on the dispersal of data fragments within a UWSN. 
The main objective of our research is based on the fact that data fragmentation, being a effective and reliable method to prevent data exploitation, must be used, even at the cost of more energy consumption.
However, to improve the security of data fragmentation itself, multi-hop fragment dispersal is needed.
However, this incurs more data transmissions in the network, increasing energy consumption of the network.
Thus, our first contribution is finding the relation of the distance of data fragments from the origin in regards to an assumed attacker model.
Also, our model analyzes the effect of remote fragmentation and its relations to energy consumption.
Then, we focus on the fact that multi-hop data fragment dispersal requires a multi-hop routing protocol, which is bound to create even more network overhead.
To solve this problem, we consider a method to transfer data fragments to remote sensors without utilization of costly existing routing protocols. 
More specifically, we propose a sink-node oriented route configuration protocol, which show how existing routing protocols can be modified to not induce excessive overhead to each sensor node, while majority of the processing is focused on the sink node. 
We compare the performance of our proposed method to other data fragment dispersion methods, and show that our method is superior to its counterparts in terms of data protection and energy efficiency. 
For our analysis we utilize simulation-based modeling tools to strengthen our initial beliefs, and then utilize the FIT/IoT-Lab experimentation environment \cite{FITIOT} to test our proposed scheme at a more practical scale.
\section{Related Work}

We categorize the related state-of-the-art to three categories. 
First, we describe the general methods of considering data survivability in the UWSN. 
Then, we emphasize the importance of data fragmentation especially for protection of the details of the data, and explain some literature in regards to this. 
Finally, we describe some possible methods of the placement of data fragments in UWSN.

\subsection{Regarding data survivability in UWSN}

In the early stages of research in UWSN, several types of methods to provide data survivability was proposed, as well as the attacker models that can be used to exploit them. 
In regards to data dispersion, there has been some methods used to disperse sensing data in an area of sensor networks. 
One of the first researches in UWSN \cite{Dipietro_2008} describe some simple methods, such as DO-NOTHING (Sensors will keep their own data), MOVE-ONCE (Each sensor will attempt to transfer its data once to another sensor node), and KEEP-MOVING (continuously move the data between sensor nodes until the sink is found). 
Some more recent work\cite{Dipietro_2011}\cite{Aliberti_2017} states that against erasure attacks, replication of data, where several copies of the same data are transferred to other sensor nodes, prove to be the most efficient. 
To provide efficient replication, these researchers propose various epidemic approaches based on controlled flooding.
%

However, while the replication of data is efficient against data erasure attacks, it becomes even more susceptible to data acquisition attacks, where the motive of the attacker is to steal and analyze the sensing data.
This is because an attacker only needs to access one of the many replicas in the network to fully grasp the information of the replicated data. 
On the other hand, data fragmentation method can effectively prevent this problem, because multiple number of fragments must be collected by an attacker to fully understand the details of one data. 
Even though it was stated that data fragmentation can be more susceptible to erasure attacks compared to data replication\cite{Dipietro_2008}, this can be somewhat alleviated if number of fragments $f_k$ is considerably higher than the required number of fragments $f_k^\prime$. 

\subsection{Data fragmentation}

Data fragmentation is another method of data management in UWSN. 
When a data is sensed, the original data $k$ is fragmented into several smaller or same-sized $f_k$ fragments, where each fragment is a variation of the original $k$ that can be individually used to decode the data. 
For an itinerant sink to successfully decode the data, it will require at least a pre-determined $f_k^\prime$ fragments, where $f_k^\prime \leq f_k$. 
This is the same case for any attacker, therefore it becomes more difficult for an attacker to successfully exploit the data and use it for its own purpose.
%

To fragment the data, several research has been presented that uses various methods such as encryption through keys\cite{Cheng_2014}\cite{Ren_2010}, and network coding\cite{Zeng_2012}.
A more recent research\cite{Kapusta_2017} states that sensing data can be efficiently fragmented based on Information Dispersal Algorithm (IDA)\cite{Krawczyk_1993}\cite{Rabin_1989}. 
In this work, the authors state that subsets of data can be fragmented through multiplications of data vectors using dispersal matrix. 
These methods operate without usage of keys, making it simpler compared to general security measures used in traditional wireless personal area networks (WPAN). 
However, methods based on IDA may have low data protection, based on the fact that specific patterns of the original data may be exploitable on the data fragments. 
To solve this problem, this research also proposes usage of stream ciphers. 
%

We note that the position of our paper focuses on the actual placement of data fragments inside a large-scale UWSN where sensor nodes can transfer data to each other via wireless multi-hop communication. 
Therefore, various kinds of data fragmentation techniques described above can be used in conjunction with our proposed scheme, regardless of their complexity. 
The performance difference between different data fragmentation methods is out of our scope in this work.

\subsection{Data fragmentation placement strategies}

Intuitively, we can consider the most basic methods for a sensor node to disperse its fragments to other sensor nodes. 
First of all, a NEAR-FIRST approach can be used, where $f_k$ fragments are directly transferred to the same number of neighbors of the originating sensor node. 
In this case, only a simple HELLO mechanism between the sensors are needed to notify each other of their existence. 
However, we will later show that dispersion of fragments to closer nodes will also increase the level of security risk. 
An alternating method to this will be the FAR-FIRST approach, where the fragments can be dispersed to sensor nodes that are furthest from the origin node.
However, this method will require well-designed multi-hop routing protocols, and sensor nodes to have explicit knowledge of the whole network architecture. 
Last of all, we can also choose a complete RANDOM method, which will also require multi-hop routing as well as usage of additional energy. 
We will not consider the aforementioned epidemic flooding methods, as they are suited more for dissemination of replicated data. 
One example of adaptive fragment placement method has been recently proposed \cite{Choi_2018}, which allows sensor nodes to place fragments to sensor nodes based on their distance and energy capacity.
This method allows data fragments to have enough distance between them, as well as prevent concentration of fragments to specific nodes to avoid energy holes and increase network lifetime.
However, this work only considers single-hop network scenarios, which is less complex to manage compared to multi-hop scenarios.
%

A problem arises in case a multi-hop routing is required, because in a traditional UWSN, a routing protocol is not needed. 
This is because the sink node travels around the network to collect the data, which allows all sensor nodes to transfer their data directly to the itinerant sink node. 
Applying traditional routing protocols in sensor networks and mobile ad hoc networks such as routing for low-power and lossy networks (RPL)\cite{RPL}, ad hoc on-demand distance vector routing (AODV)\cite{AODV}, or optimized link state routing (OLSR)\cite{OLSR}, causes additional control packet overheads. 
For example, AODV (from request and reply messages) and OLSR (from periodical topology control) methods induce heavy overhead in the entire network to find and update shortest routes between nodes. 
Also, even though RPL is considered an efficient tree-based routing method for sensor networks, it cannot be used effectively in UWSN because there is no static tree root. 
Furthermore, in the field of WSN, there has been various work that tries to create a multi-hop route between sensor nodes and a mobile sink\cite{Tunca_2014}. 
All routing methods here are out of scope in our case, because these works focus on the creation of multi-hop links between sensor nodes and the mobile sink. 
Our concern in this work, on the other hand, is creation of routes between sensor nodes themselves to share fragments. 
Therefore, all things considered, a more efficient method of routing methods are needed for efficient placement of data fragments in UWSN.
\section{Analysis of data fragmentation placement strategies}


In this section, we make a simulation-based analysis, showing that efficient placement of data fragments is needed in UWSN for more secure protection of the data.
However, also at the same time, we present some results showing that we cannot optimize the security due to energy reasons. 
For this, we design a simulation model of a UWSN environment and use it for our evaluation.

\begin{table}[t]
	\centering
	\caption{UWSN Parameters}
	\label{tab1}
	\begin{tabular}{c c c}
        \toprule
		Notation & Value & Detail\\
        \midrule
		$n$ & 100 & No. of sensors in $\sqrt n*\sqrt n$ Grid \\
		$d$ & 100m & Distance between each node \\
		$t_s$ & 600s & No. of seconds per trip for sink\\
		$v$ & 10m/s & Speed of attacker node\\
		$r$ & 20s & Time required for adversary to seize data\\
		$f_k$ & 6 & No. of fragments per data $k$\\
		$f_d$ & 3 & No. of fragments needed to decode $k$\\
		$d(f_k)$ & variant & Avg. distance between each fragment $k$\\
        \bottomrule
	\end{tabular}
\end{table}

\subsection{Environment Configuration}

First of all, a grid topology is considered with $\sqrt n*\sqrt n$ sensors, where value $n$ is configured to $100$, as shown in Table \ref{tab1}. 
In our scenario, the difference in the number of sensors will not play a big role in the difference of performance, as the effects of the size of the network and its density can be also emulated through the speed and mobility of the itinerant sink and the attacker. 
Therefore, in our case, we will statically configure the size of the network while varying the metrics of the sink and attacker instead.
In the grid, each sensor node is capable of wireless communication with all of its adjacent sensor nodes, with the distance between each vertical-horizontal node configured as $d$ meters.
Therefore, the nodes not on the edge of the network will have 4 neighbors each, with transmission range higher than $d\sqrt2$.
We select the grid topology as it is the most common deployment of sensors, and it can provide a basic understanding of the behavior of the network environment. 
Other topologies can be used to test more dynamic and extreme cases of behaviors, which differs in density of sensors (nodes concentrated in specific areas of networks), congestion of data traffic (line topologies where data are concentrated to specific nodes), and etc.. 
In our preliminary analysis, we will focus more into the basic behaviors of the network while we test a different topology in our real-life experiments shown later in our work.
%

In the network, we assume that there is one itinerant sink node $s$ which makes a trip to all the nodes in the network per every period. %
This means that $s$ will be in $d\sqrt2$ meters of each node at least once per trip. 
The time for one periodical trip is $t_s$ seconds. 
Therefore, every $t_s$ seconds, when a sensor node meets the itinerant sink, it will transfer all data fragments that it has collected in its storage, and then purge the storage so that the data cannot be retrieved by the attacker. 
%

In the case of the attacker node $a$, it is assumed to launch a data retrieval attack on the sensor nodes. 
This means that its main priority is to make a physical connection to each sensor in proximity, access the storage, and seize all data inside it. 
On the other hand, we do not consider eavesdropping attack as a method to seize data fragments. 
This is because UWSN naturally has delay-tolerant characteristics, and generation of data fragments can be irregular with high delays. 
Therefore, it is difficult for the attacker to predict when fragments are transmitted between nodes.

After a successful seizure attack, the attacker node will move its location to another sensor node, traveling at a speed of $v$ m/s. 
For the mobility model, we assume a simple Manhattan mobility model\cite{MANHATTAN}, which is suitable for modeling a random movement inside the grid environment. 
If we assume that it takes $r$ seconds for an attacker node to seize data from one sensor node, average time needed for attacking one sensor node will be $(d/v)+r$ seconds.
The number of attackers can be multiple, but should be limited to small numbers, as large-scale attacks may be discovered by the service provider. 
Taking these models into account, an attack will be considered successful if the attacker nodes acquire $f_k^\prime$ fragments before all the fragments are purged. 

Note that in our current analysis, we do not consider more sophisticated and high-performance attack models as described in previous work due to space limitations\cite{Ma_2009}. 
These sophisticated attacks can increase the probability of a successful attack, or even downright invalidate the need for multi-hop dispersion of data if an attacker can comprehend all sensor nodes even before the itinerant sink can make one successful trip.
However, we argue that this is unrealistic in the sense that this will dramatically increase the cost and complexity of the attacker, as well as become easily exploitable by the service provider so that it can take other evasive measures to protect against the attack.
In any case, our goal in this work is to analyze the danger of a realistic mobility and attack behavior of attackers (such as drones) so that it can be used as a reference to future studies, and that future works can be enhanced from this stage of work.

\subsection{Analysis of exploitation}

In the start of the network, a random sensor node will create data $k$ and $f_k$ number of fragments of $k$. 
When the fragments are dispersed to other sensor nodes, we assume that each sensor node can have at maximum $1$ fragment of that corresponding data.
This will force the attacker node to be mobile, as it must comprehend more than one sensor to be able to retrieve enough fragments to successfully decode the data. However, in some other scenarios, this value can be increased up to $f_d - 1$ per sensor node, depending on its energy capabilities or centrality.
In any case, the reason that this value must be limited to $f_d - 1$ is to prevent stationary attackers attacking only a single node, where they can exploit sensing data if more than $f_d -1$ fragments of the same data are stored at one sensor node at any time. 
%

When dispersing the fragments to other sensors, a major factor that must be considered is the average distance between the location of each fragmented data of $f_k$, which we define as $d(f_k)$. 
Therefore, a higher value of $d(f_k)$ will mean that an attack node must also travel further in average to retrieve a fragment of the same kind. 
Therefore, our first evaluation is to observe the performance of the network based on the value of $d(f_k)$ when dispersing the fragments to other sensors.
We use MATLAB software to analyze this, using the configurations of parameters shown in TABLE ~\ref{tab1}. 
The main performance we want to observe is the percentage of data seizure, which we define as the successful ratio of attack compared to the failed attacks of data seizure. 
Afterwards, we also tune other parameters such as $t_s$, $(d/v)+r$, and $f_k$ to observe their effects on the percentage of data seizure. 

\begin{figure*}[!h]
	\begin{center}
		\subfigure[Distance between fragments $(d(f_k))$ vs Data seizure percentage]{
			\label{fig1a}
			\includegraphics[width=5.8cm]{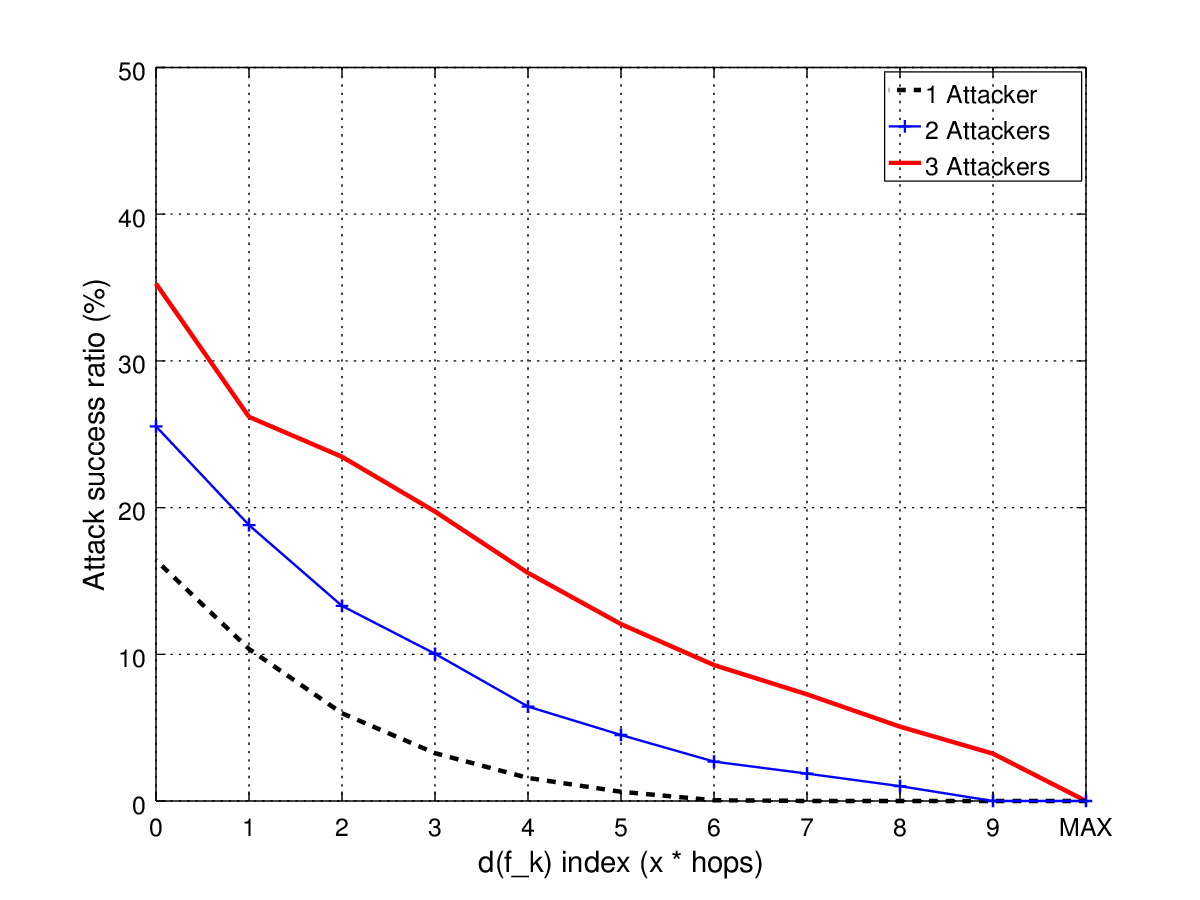}}
		\subfigure[Sink trip time $(t_s)$ vs Data seizure percentage]{
			\label{fig1b}
			\includegraphics[width=5.8cm]{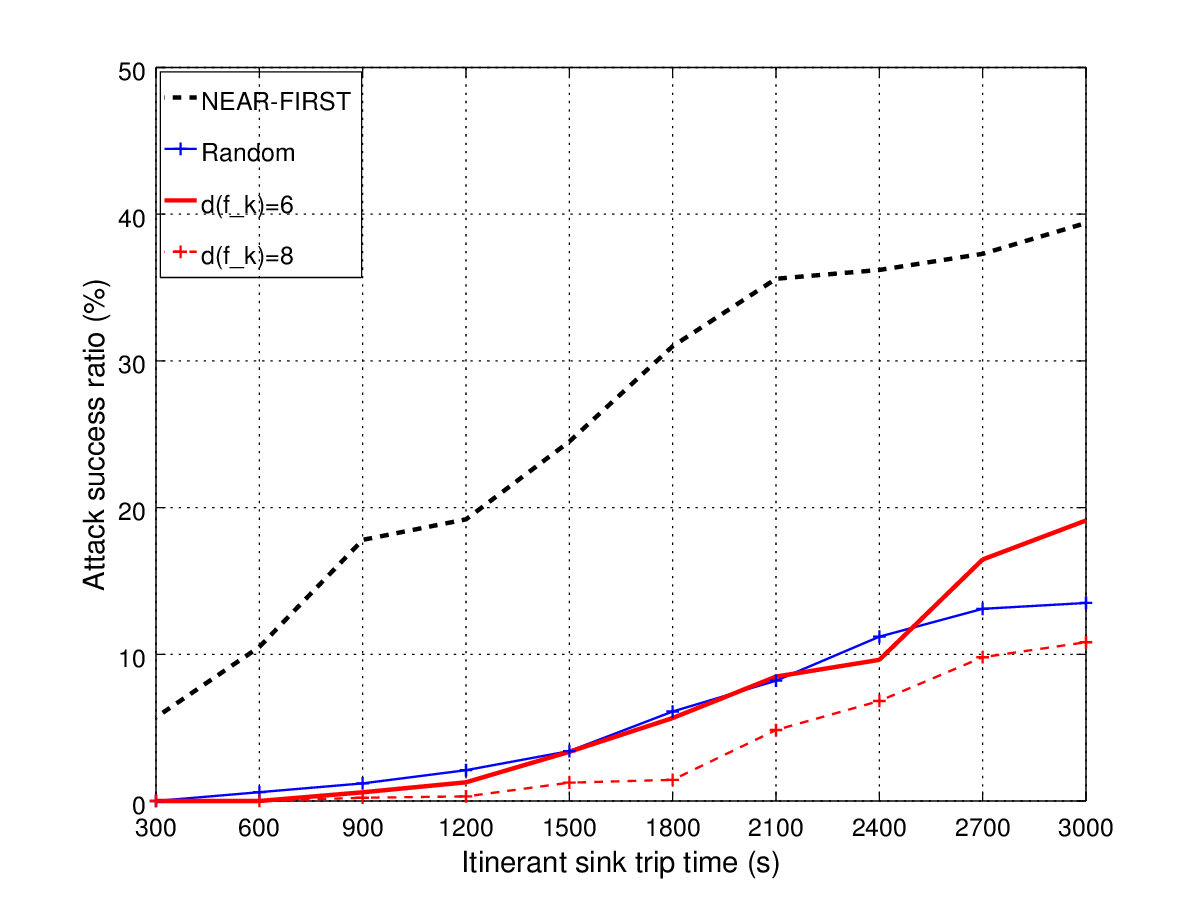}}
		\subfigure[Number of fragments $(f_k)$ vs Data seizure percentage]{
			\label{fig1c}
			\includegraphics[width=5.8cm]{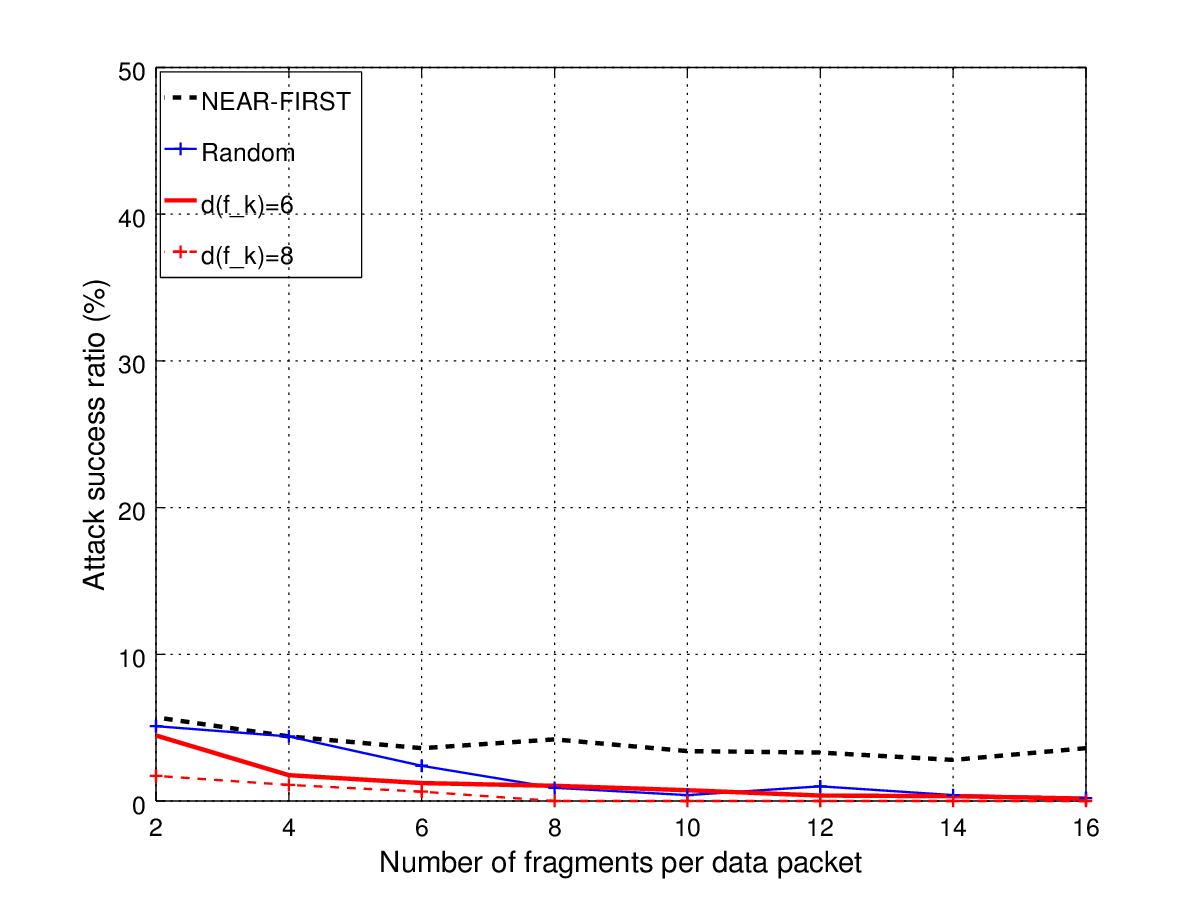}}
		
		\subfigure[Fragments required to decode $(f_d)$ vs Data seizure percentage]{
			\label{fig1d}
			\includegraphics[width=5.8cm]{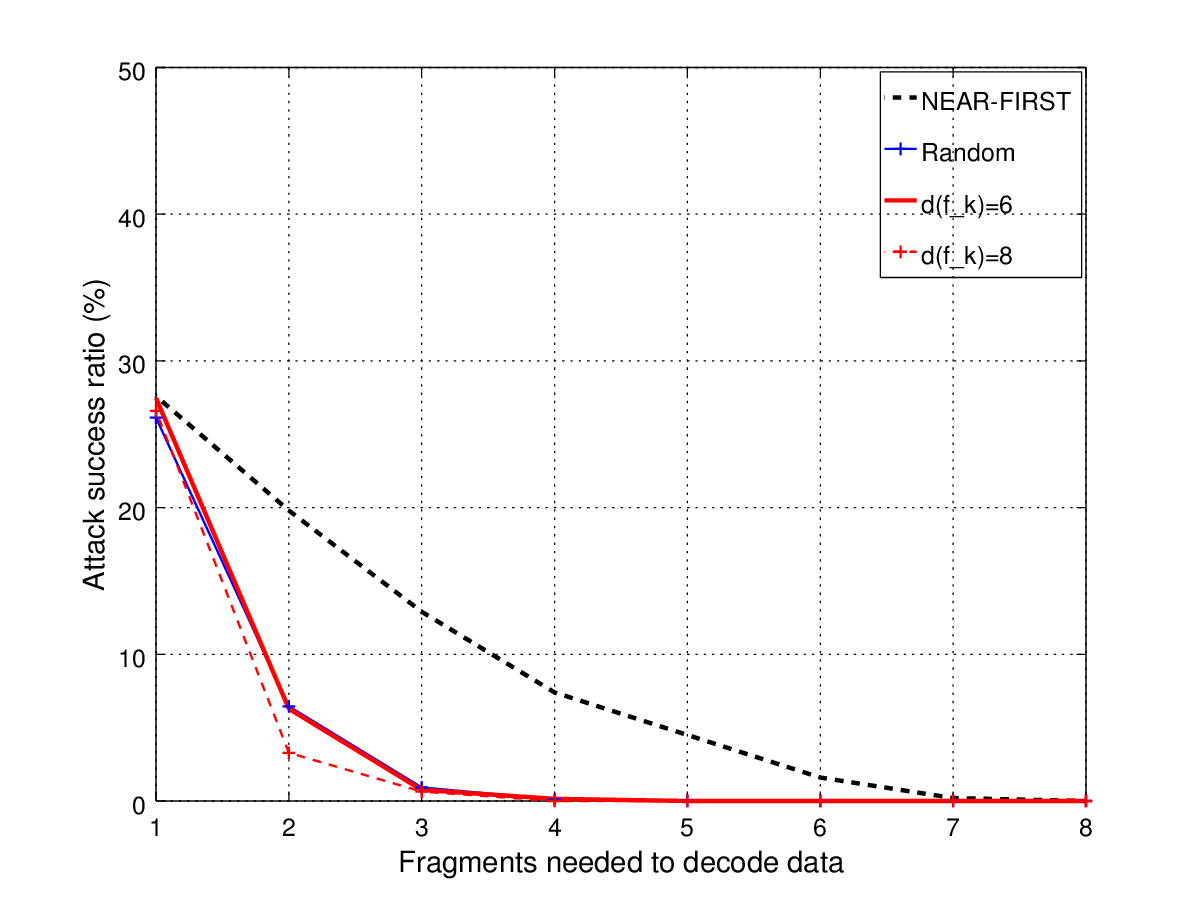}}
		\subfigure[Number of attackers vs Data seizure percentage]{
			\label{fig1e}		
			\includegraphics[width=5.8cm]{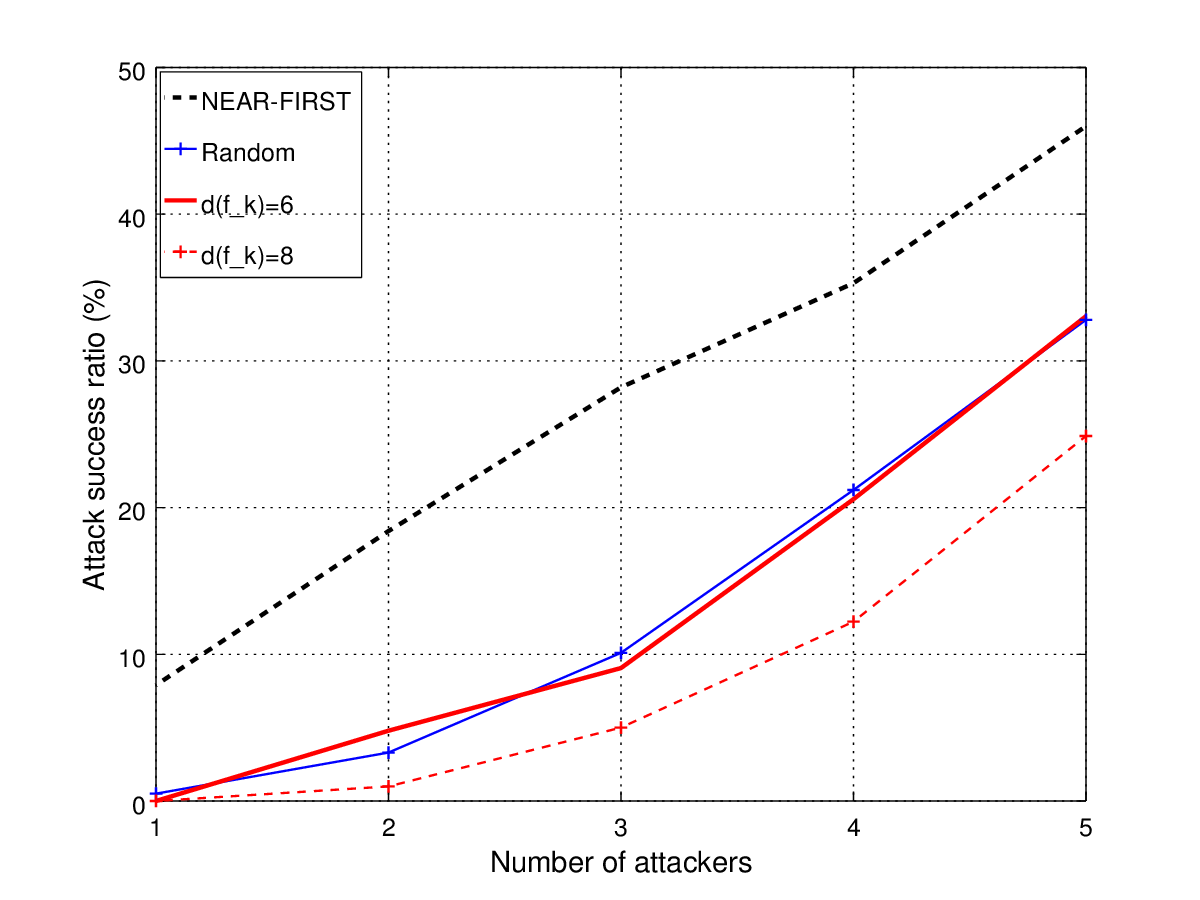}}	
		\subfigure[Distance between fragments $(d(f_k))$ vs Average $e_k$]{
			\label{fig1f}		
			\includegraphics[width=5.8cm]{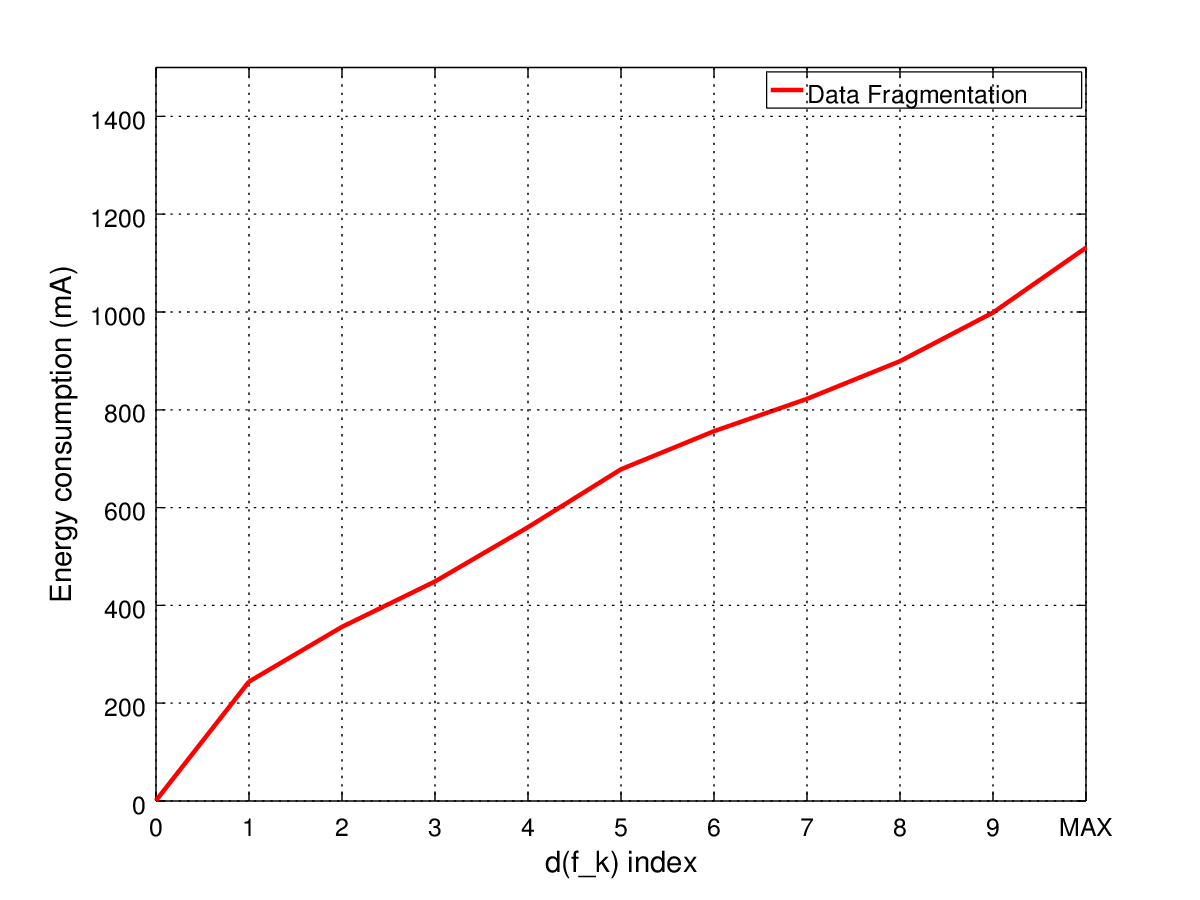}}	
		
	\end{center}
	\caption{Analysis of data fragment dispersion}
	\label{fig1}
\end{figure*}

In Fig. ~\ref{fig1a}, we measure the effect of $d(f_k)$ on the attack success ratio of the attacker. 
We denote the x-axis as an index of $d(f_k)$, which is the value denoting number of hops. 
We can clearly observe that high $d(f_k)$ value can significantly lower the success ratio of data seizure by the attacker node. 
As the attacker node is also a mobile device which may be in flight (drone) or on the move (human, vehicle), the physical distance of the fragments plays an important role on the survivability of the data.
However, it is not just the distance between the originating sensor node and the other fragments. 
It is in fact also important to scatter all fragments throughout the network in a distributed manner, to increase the average value of $d(f_k)$. 
As the number of attackers increase, we can observe that only the maximal value of $d(f_k)$ guarantees full security. 
Here, maximum $d(f_k)$ denotes the highest value of average distance between fragment, which in our scenario was near equivalent to $10$.
The maximum value can slightly vary depending on the original position of the generated data.
However, selecting the maximum value is not always the most efficient, as there are less number of choices to acquire the maximum, and the selection of sensor nodes for fragmentation becomes more easily predictable (edge/corner of the network).
Furthermore, we will also see that higher values of $d(f_k)$ result in additional energy consumption.
Therefore, it is important to find the best trade-off between these parameters suitable for each scenario. 
In the next section, we explain our method of fragment placement based on this problem.
%

Fig. ~\ref{fig1b} through Fig. ~\ref{fig1e} shows some results of basic placement techniques, NEAR-FIRST, RANDOM placement, and static $d(f_k) = 6$ and $8$, which was obtained in Fig. ~\ref{fig1a} with 0\% attack chance when there is only one attacker. 
Fig. ~\ref{fig1b} and Fig. ~\ref{fig1d} shows the attack success ratio depending on some parameters that can be controlled from the service provider's side. 
We can see that more frequent visits made by the itinerant sink can make the network more resilient. 
However, this can also increase the overall cost of the network, as well as increase the energy consumption of both the itinerant sink and sensor nodes.
Therefore, the value of $t_s$ should also be configured based on the trade-off between energy and security.
In Fig. ~\ref{fig1c}, we vary the value of $f_d$ while configuring the $f_k = 2f_d$. 
Contrary to our intuitive belief, increasing the number of fragments per data does not dramatically affect the resiliency of the network, as shown in the figure. 
This is mainly due to the fact that while increasing $f_k$ actually increases a chance that a fragment can be exploited by the attacker, increase of $f_d$ makes it also more difficult for the attacker to decode the message. 
Rather, it is in fact the percentage of $f_d$ in relation to $f_k$, that affects the resiliency of the network.
In Fig. ~\ref{fig1d}, we configure $f_k$ to 8 while the $f_d$ value is varied.
We can observe that a higher level of $f_d$ guarantees higher resiliency to attacks. 
However, if $f_d$ becomes closer to $f_k$, several other problems may occur. 
One example is that it becomes especially weak against erasure attacks, where deletion of just one fragment also prevents the itinerant sink node to successfully decode the data. 
Note that to provide a high rate of resiliency, more fragments must be used, as low number of $f_k$ cannot guarantee a high percentage of $f_d$ in relation to $f_k$. 
%

On the other hand, there are also several factors that can be controlled by the attackers. 
For example, a shorter value of $(d/v)+r$ will cause the same effect of increasing the $t_s$ value of the sink, because the attacker will have more chances to attack more nodes before the trip is concluded by the sink node. 
Also, increase in the number of attackers affect the data seizure percentage, as shown in Fig. ~\ref{fig1e}. 
Depending on the attacker's objective, the number of attackers can be increased to make the attack more efficient, but at the same time, too much increase can make the attack more revealing to the service provider. 
We can observe through these results that multi-hop placement of data fragments is an important technique that needs to be addressed for better management of a UWSN. 

\subsection{Analysis of additional overhead}

Even though multi-hop dispersion of fragments can increase data survivability, it will also induce additional energy usage. 
This is quite intuitive as seen in various communication modules used in sensor networks\cite{AT86}, where every time spent in transmission ($t_{tx}$) and reception ($t_{rx}$) causes a device to use more energy than being in an IDLE state. 
For example, per each fragment created by a sensor node, the resulting additional energy overhead $e_f$ per fragment can be calculated as

\begin{equation}
e_f = \sum_{i=1}^{h_f} (e_{tx}*p_i + e_{rx}),
\label{energy_frag}
\end{equation}

where $p_i$ denotes the percentage of the maximum transmission power ($0 \leq p_i \leq 1$) of each transmitting node in the multi-hop transmission, and $h_f$ represents the number of hops needed to deliver the fragment to a destination's sensor node. 
Note that we do not consider any losses and retransmissions due to instability of the network. 
Therefore, energy consumption induced by each data $e_k$ generated by a sensor is

\begin{equation}
e_k = \sum_{i=1}^{f_k} e_f,
\label{energy_data}
\end{equation}
%

On the other hand, dispersion of data fragments in single-hop scenario will only induce $e_k = \sum_{i=1}^{f_k} (t_{tx}*p_n + t_{rx})$. 
Therefore, intuitively we can see that higher value of $h_f$ will cause steadily increase in the energy consumption. 
We visualize this equation through Fig. ~\ref{fig1f}, which shows the average amount of energy used in the network per data based on an actual energy model defined in \cite{AT86}.
We can observe that the increase in energy consumption is quite linear, as the increase in number of hops is also linear to value of $d(f_k)$. 
As mentioned previously, selecting the maximum value of $d(f_k)$ is not efficient because the location of of fragments become easily predictable. 
Therefore, at least in our scenario, a value of $d(f_k) = 6$ or $8$ will be most efficient performance wise, which can be decided depending on the number of attackers in the network.
Note that different $d(f_k)$ values will be efficient in different cases of network, and we show that these values can be found through our method of analysis.
%

So far, we have observed the energy consumption generated from additional transmission induced from fragments. 
However, another major problem here is that we have only considered so far the additional transmission overhead, without consideration of the routing process itself. 
Generally, to create the routing tables to transmit data in a multi-hop fashion, additional control packet overhead is also induced by the routing protocol. 
First of all, we consider periodical transmission of HELLO message. 
This is the most basic transmission by each node to notify their existence in the network/routing layer. 
Therefore, this can be considered an inevitable overhead that exists in most networks. 
However, other popular routing protocols are already known to generate various control overheads. 
AODV\cite{AODV} is prone to generating much control overhead, especially during the route request process in UWSN. 
This is because the reactive properties of AODV does not create a total map of the network, and each node wanting to send data must flood request packets to the entire network every time a destination needs to be found. 
On the other hand, OLSR\cite{OLSR} fares better in this scenario because a global routing map can be created proactively for every sensor node. 
However, the proactive route update for topology control in turn causes periodical broadcasting in the network, which is also a heavy process. 
To remove frequent broadcast or flooding in the network, geographical routing methods such as Greedy Perimeter Stateless Routing (GPSR)\cite{GPSR} can be used, which utilizes a greedy data forwarding mechanism without any proactive or reactive route table updates. 
However, GPSR needs location information for each sensor, and usage of global positioning system on sensor nodes in turn uses energy and unsuitable for small, embedded sensor devices.
%

In our main proposed scheme, we consider how we can reduce energy consumption of the routing protocol in UWSN. 
As we have already explained how existing routing protocols in the mobile ad hoc networks cannot be used in this case, we make modifications to make them suitable for dispersion of data fragments in UWSN. 

\section{Multi-hop Data fragmentation}

Our main goal of the proposed scheme is to develop a routing protocol suitable for UWSN, which uses only HELLO messages to share between sensor nodes and the itinerant sink to create routes for fragment distribution in UWSN. 
The main idea of the scheme is to: (a) minimize control packet transmission between sensor nodes for creating routes, and (b) let the itinerant sink node to do most of the job. 
The scheme can be divided into four phases: (1) HELLO transmission, (2) Data collection and storage, (3) Route creation, and (4) Route distribution.

\subsection{HELLO transmission}

Each sensor periodically transmits a HELLO packet with a predetermined transmission power and time. 
For simplicity, we will not take into account power control schemes that may affect the number of neighbors for each sensor node. 
Inside each HELLO packet, a sensor node includes its ID, ID list of its neighbors (which is obtained through the HELLO packet itself), and the expected transmission count(ETX)\cite{Couto_2004} of each neighboring node's HELLO packet when they are received. 
The ETX will be used to judge the transmission quality of the link between each neighboring node in the routing layer, by counting the number of HELLO packets actually received compared to how much that should have been received over a period of time. 
This also allows the itinerant sink node to analyze if each link is bi-directional or unidirectional. 
For the link quality measurement, other information from the physical layer such as received signal strength indicator (RSSI) or signal to noise ratio (SNR) can be used. 
However, it is also known that these physical layer signal measurements can be affected by deteriorated environments and may not be accurate\cite{Parameswaran_2009}\cite{Heurtefeux_2012}. 
Therefore, in our case, we utilize a routing layer information which explicitly shows direct information on the successful transmission of HELLO packets.

\subsection{Data collection and storage}

\begin{figure} [t]
	\begin{center}
		\includegraphics[width=8cm]{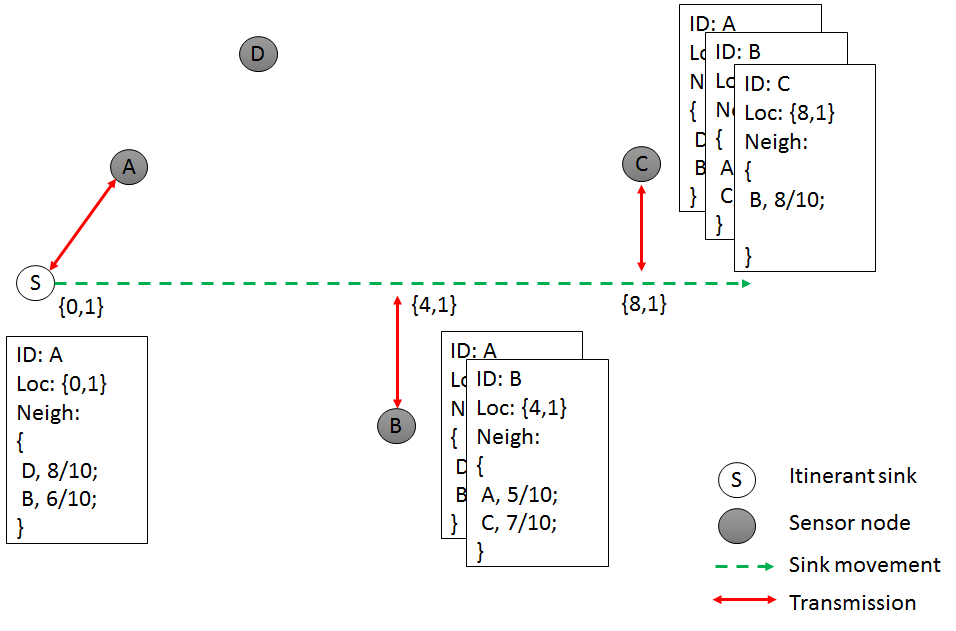}
	\end{center}
	\caption{Data collection and storage by itinerant sink}
	\label{fig2}
\end{figure}

While the HELLO packet is constantly generated, shared, and updated by all the sensor nodes in the network, itinerant sink will also listen to the HELLO of all the sensor nodes while making its pre-determined trip following a mobility model. 
We assume that the itinerant sink will be in the communication range of all sensor nodes at least once while on the move. 
Therefore, the itinerant sink can obtain the neighboring node information of all nodes, thus creating a list for each node $n_s$ containing node ID, node's neighbor list and link quality $neigh_{n, ETX}$. 
Also, a rough indication of each node's location is recorded, by recording the itinerant sink's own coordinates when a HELLO is received while on the move. 
For this, a positioning system needs to be installed in the drone, which is a common practice in recent applications\cite{Kerns_2014}. 
An example of data collection by the itinerant sink and the format of data storage is shown in Fig. ~\ref{fig2}.

After gathering lists from each node in the network, the itinerant sink can create a global graph of nodes and edges based on the neighbor list. 
This graph will result in a bi-directional link multi-hop network, which will be used by the itinerant sink to generate routes using routing protocols.

\subsection{Route creation}

After the itinerant sink makes a full trip of the entire network, the sink will commence the selection of candidates that will store the fragments and multi-hop paths for each of these nodes. 
Firstly, based on the graph created by the itinerant sink, it can create a crude map of the entire network. 
Using this information, the sink will cluster all sensor nodes into areas recorded through its location. 
The number of clusters can be similar to a pre-determined $f_k$ for all sensor nodes, which would allow one data fragment per cluster. 
For example, if $f_k = 4$, the entire network will be divided into four even areas, and then each sensor will be classified into each area depending on the location it was recognized by the sink node. 
One of the most simple methods of clustering the sensors is to use a $k$-means clustering\cite{Bock_2007} algorithm, which accepts the ${x,y}$ coordinates as the criteria to categorize while $k = f_k$. 

As the itinerant sink is making a continuous trip through a pre-defined mobility model, there is a very high chance that a sensor can be recognized by the sink node in more than one location. %
In this case, before the $k$-means clustering phase, the multiple coordinates of each node can be merged to find the average point, which would actually denote a more accurate location information than using only one coordinate. 
For example, if three or more points are found per node, then tri- or multi-lateration \cite{Madeni_2010}\cite{Villas_2013} can be used to tune the location of each sensor node through the RSS information received by the sink and the location of where each sensor node was sensed.
%

After each sensor node is clustered into areas, one sensor node is randomly selected per each area. 
Therefore, there will be $f_k - 1$ sensor nodes that are selected for each sensor node's destination for data fragmentation, excluding the fragment that will remain at the originator sensor node. 
Note that for our current work, we select a simple random selection method for selecting fragment destinations. 
This will in fact not guarantee the highest $d(f_k)$ for all fragments. 
However, if we always try to achieve the highest $d(f_k)$ for destinations, the fragments will always be transferred to sensor nodes that are at the edge of the network, causing unbalanced dispersion of fragments within the network. 
This makes the fragmentation predictable and causes these edge nodes to become higher priority nodes to be selected for an attack if the attackers are aware of the algorithm. 
Furthermore, the edge nodes and the intermediate routing nodes towards the edge nodes will consume much more energy, which will result in reduction of network lifetime. 
%

In some cases, complete random selection of sensor nodes in each cluster may pose some problems; for example selection of two neighboring sensor nodes in different clusters. 
However, this can be easily avoided, as the itinerant sink can compare the locations of each sensor node upon selection and re-select if such case occurs.
This is possible as the itinerant sink has global information of the network.
%

After random selection of sensor nodes, for each sensor node $n_s$, sink node selects one multi-hop route for each selected destination $n_d$, eventually creating $f_k$ multi-hop routes in the process. 
To do this, a specific multi-hop wireless routing protocol must be used. 
Here, we show three possibilities of using existing routing techniques modified to work in our case: 1) Source routing, 2) Table-driven routing, and 3) Geographical routing.

\subsubsection{Source routing}

As an simple example, source-routing such as Dynamic Source Routing (DSR) protocol\cite{DSR} can be used, where multi-hop routes are normally created using route request broadcast packets and route reply packets.
This is possible because the sink node maintains a bi-directional graph of all nodes in the network, based on the lists of neighbor information.
Therefore, instead of running the DSR protocol in each of the nodes via a distributed manner, we compute it all at the sink node via a centralized manner. 
The main difference is that for the routing metric, we utilize the ETX, which is a better choice for bi-directional link compared to the generic hop-count metric. 
DSR's route request process is heavy in communication point of view as it requires flooding of messages, but can be a much simpler process if done only in a computational manner by the sink node. 
Algorithm \ref{alg1} shows the process of DSR inside the itinerant sink.

\begin{algorithm}
  \caption{Centralized DSR}
  \label{alg1}
  \begin{algorithmic}[1]
	\Function{RouteCreate}{$src$, $neigh_{src}$, $dest$}
      \For {each $n$ in neighbor list $neigh_{src}$}
          \State{$met = n_{etx}$}
          \State{$seq = src$}
          \State {RouteRequest($n$, $met$, $seq$, $dest$)}
      \EndFor
	\EndFunction
  \end{algorithmic}

  \begin{algorithmic}[1]
	\Function{RouteRequest}{$int$, $met$, $seq$, $dest$} 
        \If {$int \in$ $dest$}
          \State{STORE $seq$ and $met$}
          \State{SELECT $seq$ with lowest $met$}
  	      \Return{$seq$}
  	    \EndIf
        \If {already called RouteRequest AND $met$ is higher than stored $met$}
        
          \Return{}
  	    \EndIf

      \For {each $n$ in neighbor list $neigh_{int}$}
      \State{$met += n_{etx}$}
      \State{$seq = seq^\frown int$}
      \State {RouteRequest($n$, $met$, $seq$, $dest$)}
      \EndFor
	\EndFunction
  \end{algorithmic}

\end{algorithm}

The computational complexity of DSR in the itinerant sink will therefore be $O(n^2)$, as the \textit{RouteCreate} function will be called for each sensor node while \textit{RouteRequest} function will also be called for all nodes at most once every time a \textit{RouteCreate} function is called. 
The advantage is that \textit{RouteCreate} does not have to be called for each destination, as one flooding process is enough to discover all destinations. 
Also, the centralized DSR does not require a route reply phase, which is used in its traditional version. 
This is because the sequence of routes can be created during the route request phase and are all stored inside a global storage.
%

However, DSR also has its own drawbacks when used as a routing protocol. 
The main problem is that the multi-hop routing information is stored inside the data packet itself. 
Therefore, when a sensor node transmits its data fragment to the next-hop sensor node, it includes the entire multi-hop sequence in the packet so that this sequence can be referenced by all intermediate nodes in the sequence. 
Therefore, the additional communication overhead per packet is $S_a * N_h$, where $S_a$ denotes the size of node address and $N_h$ denotes the number of hops that packet must travel. 
The overhead can dramatically increase if the network size and number of hops increase. 
However, unlike table-driven routing methods, DSR does not incur additional storage overhead for each sensor node. 

\subsubsection{Table-driven routing}

We can also utilize AODV \cite{AODV} for utilization of table-driven routing in a centralized manner within the itinerant sink. 
In this case, the route discovery process is quite similar to DSR, which is ideal in this case as all the process of flooding route request packets are computed only by the sink node. 
However, in case of AODV, the sink node must compute a routing table for each sensor node in the network, and distribute this routing table during the next trip. 
The routing table contains the list of all destination sensor nodes where the fragments must be transmitted to, followed by the intermediate node addresses for each destination node as the next-hop forwarding node. 
Algorithm \ref{alg2} shows the process of AODV inside the itinerant sink.

\begin{algorithm}
  \caption{Centralized AODV}
  \label{alg2}
  \begin{algorithmic}[1]
	\Function{RouteCreate}{$src$,$neigh_{src}$,$dest$}
      \For {each $n$ in neighbor list $neigh_{src}$}
          \State{$met = n_{etx}$}
          \State {RouteRequest($src$,$n$,$src$,$met$,$dest$)} 
      \EndFor
	\EndFunction
  \end{algorithmic}

  \begin{algorithmic}[1]
	\Function{RouteRequest}{$src$,$int$,$prev$,$met$,$dest$} 
        \State{INSERT to table $int$: previous hop $prev$, source $src$, metric $met$}
        \If {$int \in$ $dest$}
          \State{SELECT $src$, $prev$ with lowest $met$}
  	      \State {RouteReply($src$,$int$,$int$,$prev$)}
  	    \EndIf
        \If {already called RouteRequest AND $met$ is higher than stored $met$}
        
          \Return{}
  	    \EndIf

      \For {each $n$ in neighbor list $neigh_{int}$}
      \State{$met += n_{etx}$}
      \State {RouteRequest($src$,$n$,$int$,$met$,$dest$)}
      \EndFor
	\EndFunction
  \end{algorithmic}

  \begin{algorithmic}[1]
	\Function{RouteReply}{$src$,$dest^i$,$next$,$int$} 
        \State{INSERT to table $int$: next hop $next$, destination $d^i$}
        \If {$src = int$}
          \Return{}          
        \Else
          \State {READ from table $int$: previous hop $prev$ from source $src$}
  	      \State {RouteReply($src$,$d^i$,$prev$,$int$)}
  	    \EndIf
	\EndFunction
  \end{algorithmic}
\end{algorithm}

The computational complexity of AODV is similar to DSR even though there is an extra route reply phase. 
However, the route reply phase in AODV is a unicast phase, and the overall overhead that is incurred from reply is negligible compared to the expensive flooding phase that occurs from the route request phase. 
The main problem of utilizing AODV in UWSN is that other than the destinations to sensor nodes for transmitting fragments, intermediate nodes within a multi-hop route must also record forwarding tables for other nodes that may utilize it for data forwarding. 
Therefore, if the number of sensor nodes increase in the network, so does the possibility of the routing table size, especially for sensor nodes situated near the middle of the network. 
The storage overhead for each sensor node will be $(S_a * 2)* (f_k-1) + (S_a * 2 )*N_i $, where the routing table must contain the destination address and intermediate address for each outward fragment, plus additional number of unpredictable $N_i$ intermediate routes that it may be used as a forwarding node. 
As sensors generally lack in memory size due to its constraints, this problem can limit utilization of AODV in some scenarios. 

\subsubsection{Geographical routing}

In our current UWSN environment, we consider utilization of geographical routing, more specifically GPSR \cite{GPSR}, as an efficient routing method. 
Even though we have mentioned previously that GPSR cannot be used without location information, in our case we have utilized the itinerant sink to generate a crude map of sensor nodes based on its own location. 
Therefore, the graph updated by the sink can be used in our scenario.
%

GPSR is based on greedy forwarding mechanism, where multi-hop routes are generally selected upon the physical closeness to the destination. 
The itinerant sink, using GPSR, will create a routing table for each sensor node, where each destination node's coordinates are recorded. %
When the routing table is transferred to a sensor node, the sensor node will look up the destination node's coordinates, find the sensor node in its neighbor table which is closest to the destination, and then simply forward the data to the closest neighbor. 
Algorithm \ref{alg3} shows the process of GPSR inside the itinerant sink.

\begin{algorithm}
  \caption{Centralized GPSR}
  \label{alg3}
  \begin{algorithmic}[1]
        \Function{FindCoordinate}{$src$,$dest$}
        \For{each $dest^i$ of $src$}
        
        \Return $dest^i_x, dest^i_y$
        \EndFor
        \EndFunction
  \end{algorithmic}

\end{algorithm}

As observed in this algorithm, the workload of the itinerant sink is clearly lower than that of centralized AODV and DSR. 
An iterative process is not required to search for a specific destination, as the forwarding decisions will be made by each sensor node in a greedy manner. 
This allows the process to be finished in $O(n*m)$, where $n$ equals the number of sensor nodes and $m$ equals the number of destination nodes per sensor node, which is $f_k - 1$. 
Furthermore, there are no additional communication overhead. 
In case of storage overhead, if we assume a 2-dimensional deployment, the size of routing tables are $(S_c * 2)*(f_k - 1)$, where $S_c$ denotes the size of coordinates. 

\subsection{Route distribution}

After the creation of multi-hop routes for all nodes, the itinerant sink can now begin the next trip, in which the routes are distributed to all the sensor nodes. 
Through the map provided by the itinerant sink node, a sensor node can now transmit its data fragment to the determined destination nodes. 
%

In regards to maintaining shortest routes, the centralized AODV and DSR will guarantee the shortest routes to each destination based on the best ETX value. 
However, for GPSR, the selected multi-hop routes may not be optimal as intermediate nodes may sometimes not be able to to greedy forwarding due to holes in the map (e.g. destination is not my neighbor yet I am the closest to it). 
In this case, GPSR will initiate perimeter routing, which can also be used by the sensor nodes in UWSN. 
However, we believe that if these forwarding holes are analyzed before by the itinerant sink, we can further prevent this problem. 
On how to detect map holes for GPSR in UWSN will be discussed in our future work.
\section{Performance Evaluation}

The performance of our proposed data fragmentation scheme is evaluated in the FIT/IoT-Lab environment\cite{FITIOT}. 
FIT/IoT-Lab is a large-scale infrastructure of sensors and embedded devices connected with each other through wireless communication. 
It provides various locations of sensor testbed environments in Europe, each location with a unique deployment of sensors. 
For our experiments, we specifically select the maps in $Grenoble$ site and the $Lille$ site, where we randomly select 50 sensors for each experiment. 
Each map has distinctive properties in their topology, where $Grenoble$ map has sensors installed following long corridors of indoor buildings, creating long multi-hop scenarios, while $Lille$ map has uniform grid topology of sensors in 3-dimensional space. 
Therefore, in the $Grenoble$ map, the attacker will also follow this straight line to attack sensor nodes in order, while in the $Lille$ map, the attacker will attempt a circular movement to account for all nodes in the network.
In the current version of the FIT/IoT-Lab, mobile sinks cannot be easily used. 
Therefore, we make an abstract model of the itinerant sink and the attacker, which can be implemented through a front-end management module provided by the service. 
For example, according to a mobility model, we can command a sensor node to behave as if the data was collected or it was attacked, and the information of the attack is propagated to the front-end computers for analysis. 

\begin{table}[t]
	\centering
	\caption{UWSN Parameters on FIT/IoT-Lab}
	\label{tab2}
	\begin{tabular}{c c c}
        \toprule
        Notation & Value & Detail\\
        \midrule
        $n$ & 50 & No. of sensors \\
		$f_k$ & 6 & No. of fragments per data $k$\\
		$f_d$ & 2-5 & No. of fragments needed for decoding $k$\\
        \bottomrule
    \end{tabular}
\end{table}

\begin{figure*}[!h]
	\begin{center}
		\subfigure[Dispersion methods vs Attack success percentage]{
			\label{fig3a}
			\includegraphics[width=5.8cm]{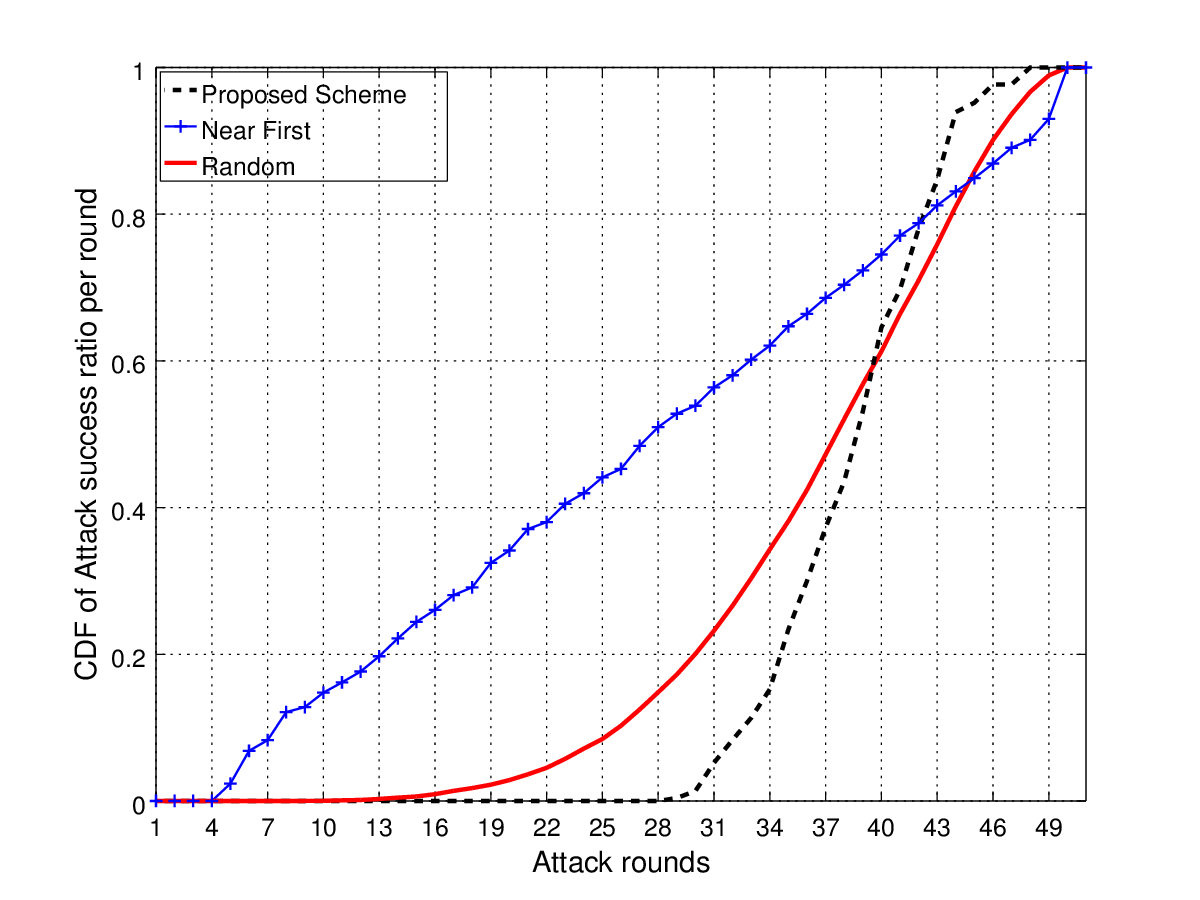}}
		\subfigure[2 attackers]{
			\label{fig3b}
			\includegraphics[width=5.8cm]{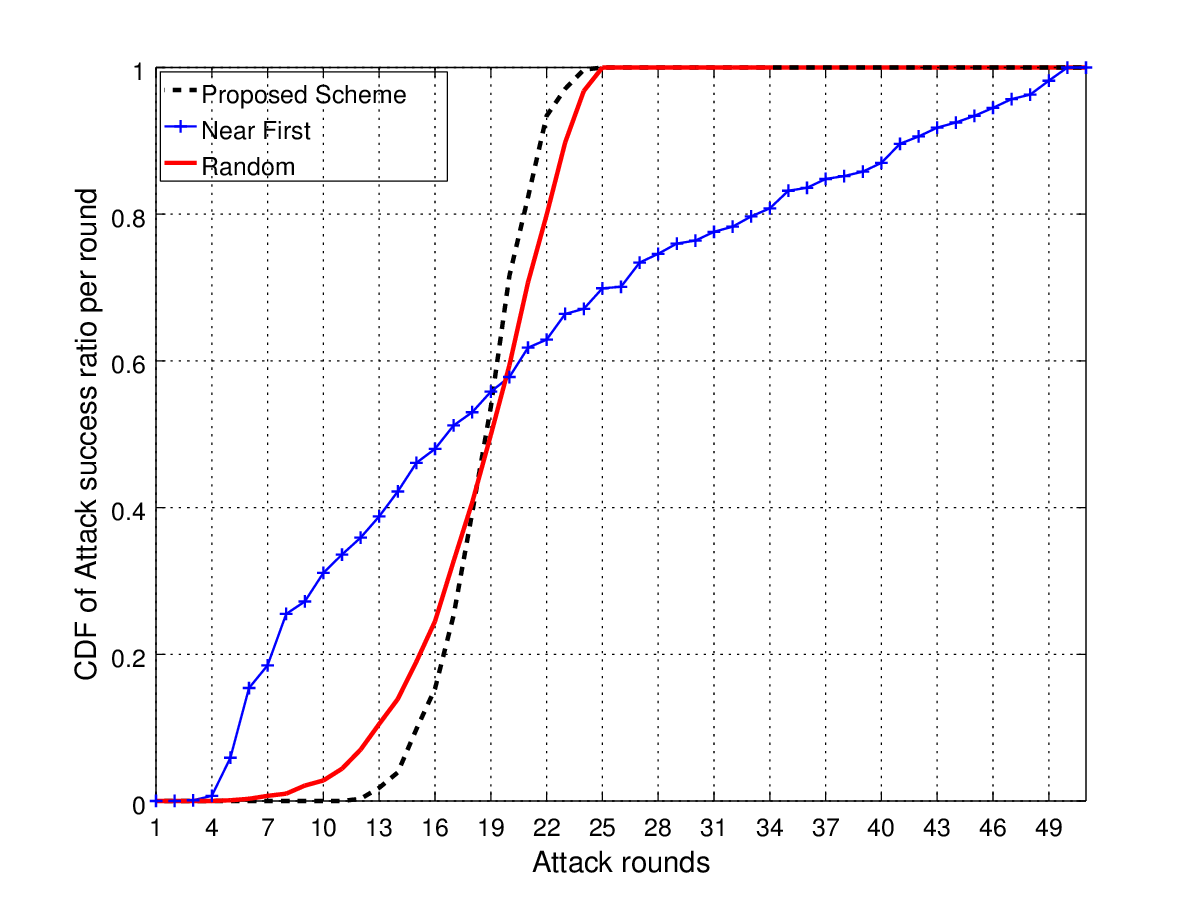}}
		\subfigure[Effect of $f_d$]{
			\label{fig3c}
			\includegraphics[width=5.8cm]{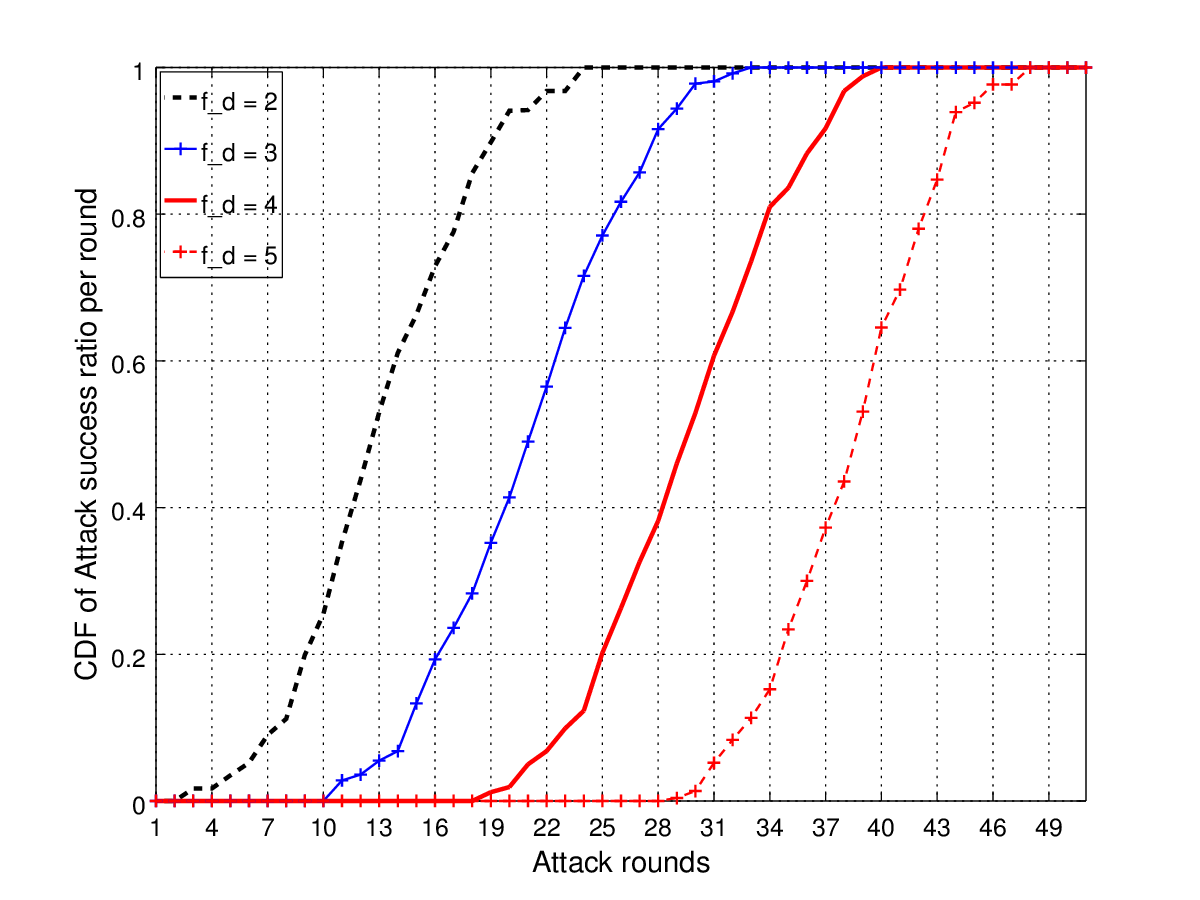}}

	\end{center}
	\caption{Performance Evaluation of Proposed Scheme in Grenoble map}
	\label{fig3}
\end{figure*}

\subsection{Security evaluation of proposed scheme}

For performance comparison, we compare our proposed multi-hop fragment distribution scheme using centralized routing techniques to their traditional routing methods, and also the random and NEAR-FIRST approach.
The main factors for comparison is the attack success ratio by the attacker. 
For the variable parameter we use the number of rounds required by the attacker to successfully exploit the data.
This means that the attacker will need to attack certain amount of sensor nodes, while the mobile sink must collect the required data before the needed number of rounds by the adversary.
In each round, one attacker will make a movement towards a node, and exploit its data.
Therefore, in this scenario, it will clearly show that more efficient fragment dispersion methods will force the attacker to use more rounds to exploit the data.
The static parameters are shown in TABLE ~\ref{tab2}.
In Fig. ~\ref{fig3a}, we observe the performance of the proposed scheme, which uses $k-means$ clustering to create clusters in the network and disperse fragments in each cluster.
This allows fragments to be fairly evenly distributed, providing the lowest attack success among the three methods.
The reason that $Near-First$ method performs better than the other methods at the ending phase (round 40-50) is due to some cases where the movement of an attacker may arrive at the location of the bunched fragments only at the very end.
However, this depends mainly on a random variable, and most security-based use cases will need a much higher requirement , for example $0\%$ attack rate, depending on the application requirements.
In case of the proposed scheme, the performance of remaining at $0\%$ attack rate is more than $100\%$ better than random method, and over 6 times higher than near-first scheme.

In Fig. ~\ref{fig3b}, we can observe the effect of having multiple attackers in the network.
Here in this scenario, we configure one more attacker which moves in the opposite direction of the other attacker, towards the center of the network.
As seen from the graph, the performance radically decreases compared to having just one attacker.
This is especially due to the topology of the network, as the limited directions of mobility allows the attackers to quickly converge towards themselves, exploiting all nodes without any redundancy in their attack.
However, even in this case, the proposed scheme performs as twice as better than the random (6 rounds vs 12 rounds).
In Fig. ~\ref{fig3c}, we observe the performance of the proposed scheme while configuring the value of $f_d$. 
Although it is quite intuitive that the more number of fragments required to decode the data provides more security, it is also true that sink node will also require more time to decode the data as well.
Furthermore, if the adversary is launching a data deletion attack, then the performance of this graph is exactly inverted, meaning that $f_d = 2$ will perform like $f_d = 5$ and vice versa.
This is because in the case of data deletion attack, for example, the adversary only need to delete $2$ fragments if $f_d = 5$.
Therefore, depending on the capability of the sink node and type of attack, it will be important to select an appropriate value of $f_d$.

\begin{figure*}[!h]
	\begin{center}
		\subfigure[Dispersion methods vs Attack success percentage]{
			\label{fig4a}
			\includegraphics[width=5.8cm]{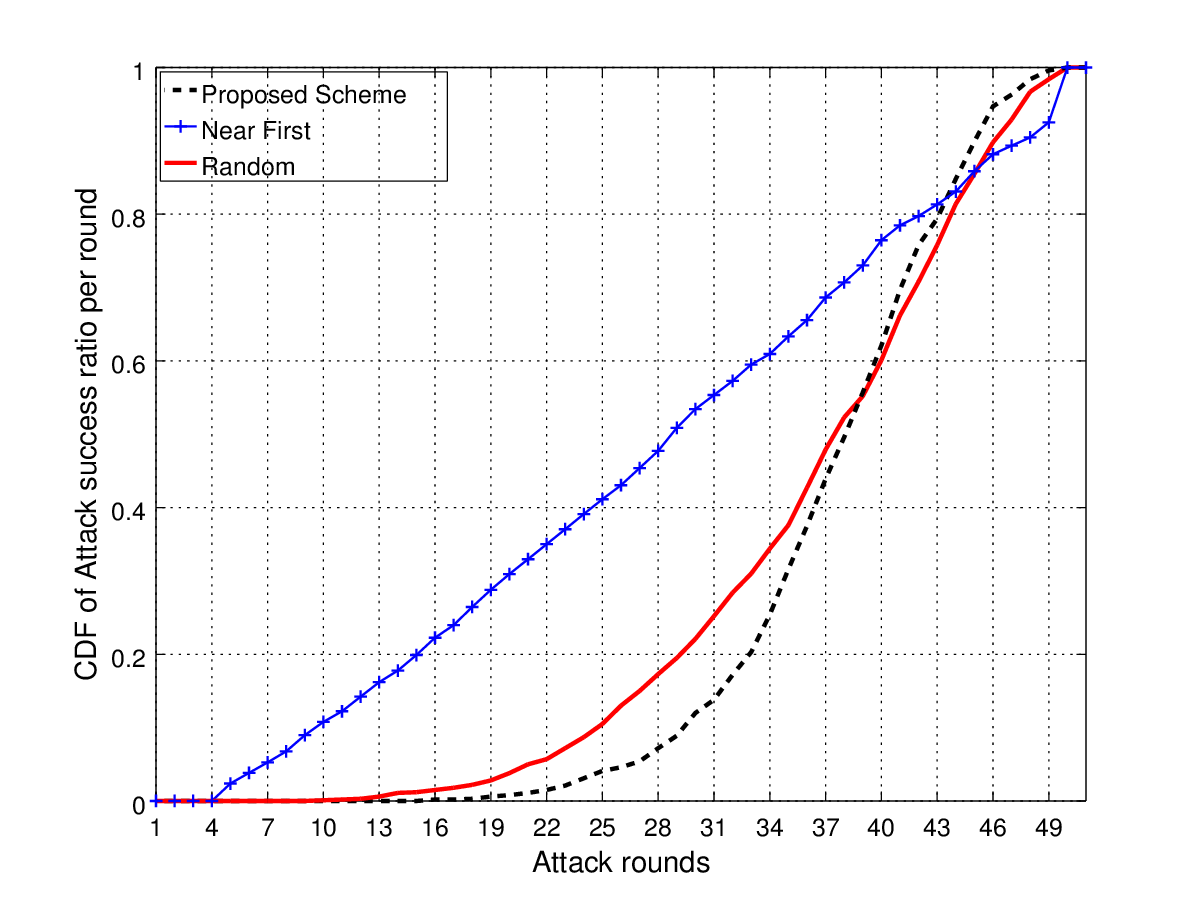}}
		\subfigure[2 attackers]{
			\label{fig4b}		
			\includegraphics[width=5.8cm]{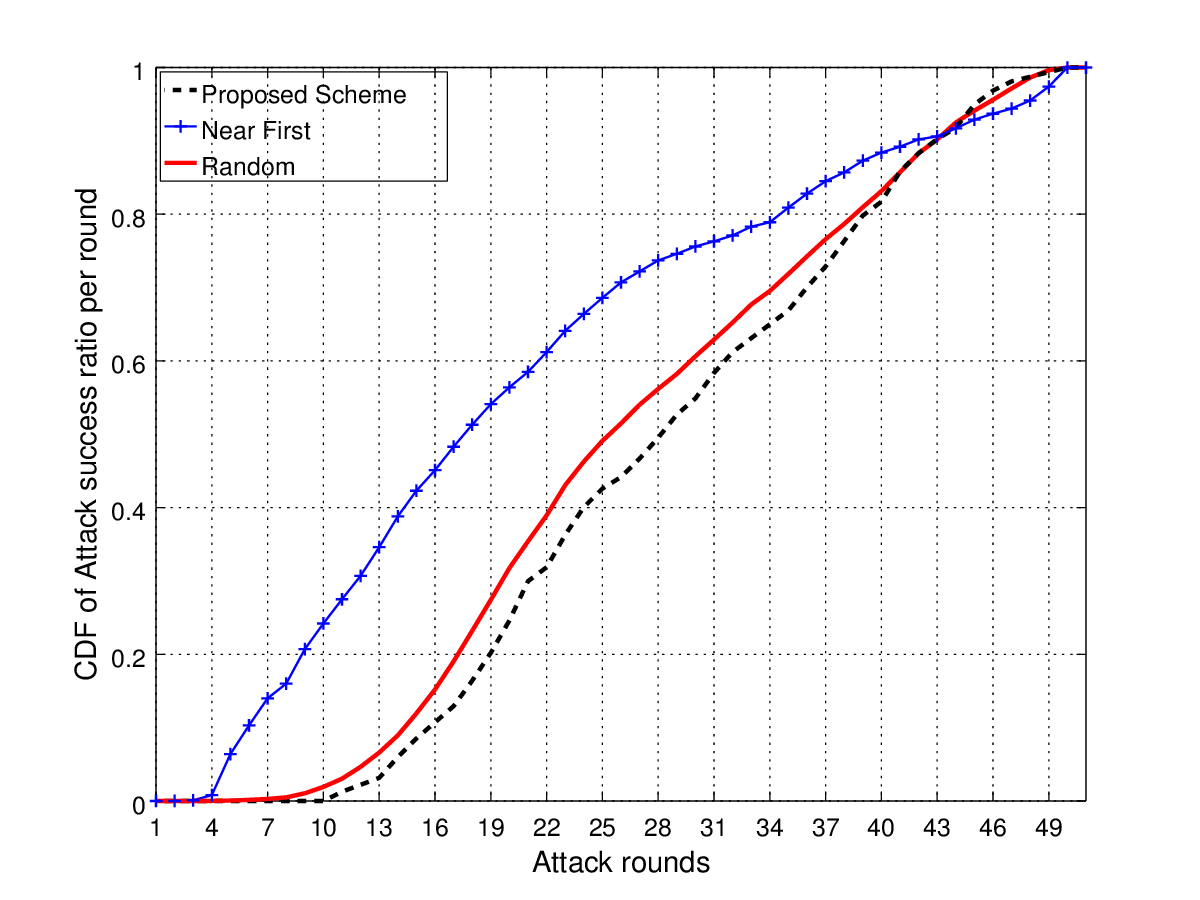}}	
		\subfigure[Effect of $f_d$]{
			\label{fig4c}		
			\includegraphics[width=5.8cm]{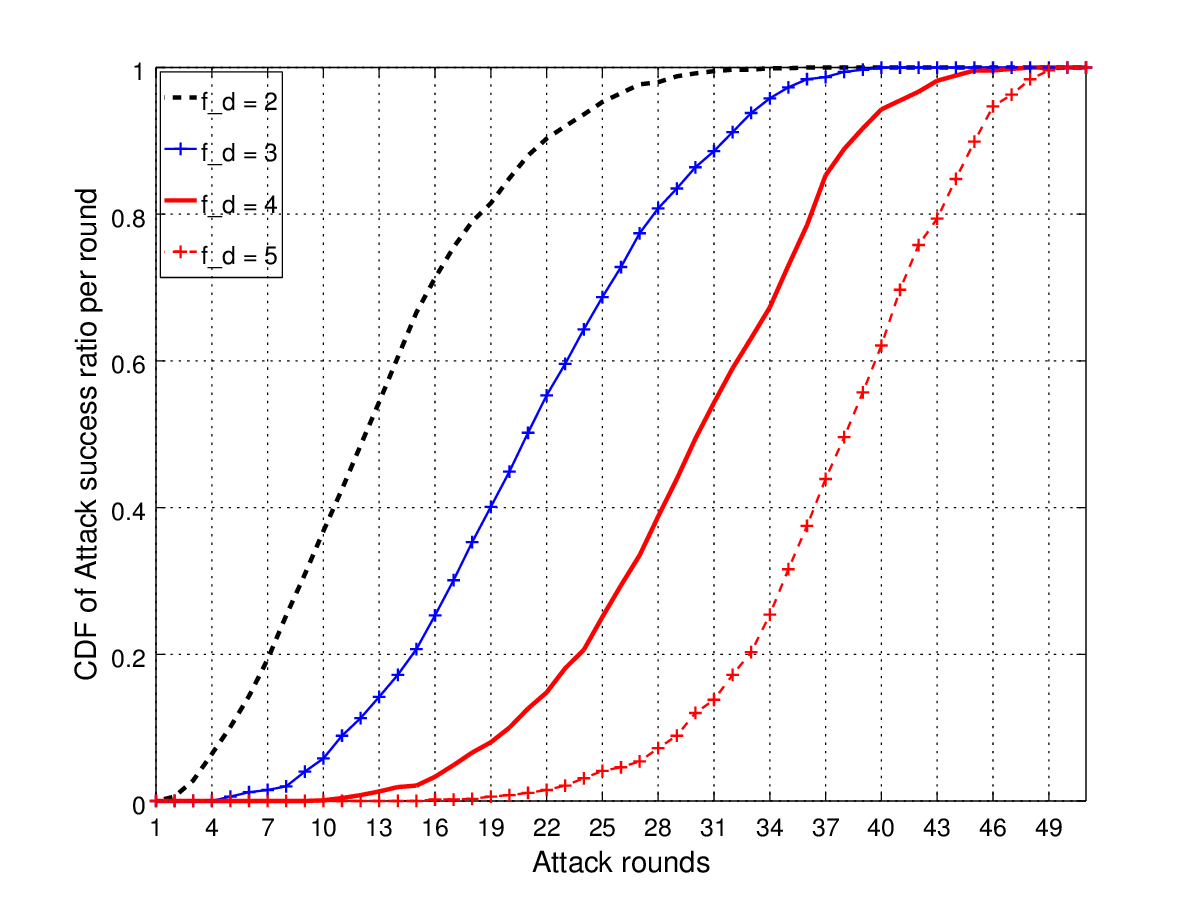}}	
		
	\end{center}
	\caption{Performance Evaluation of Proposed Scheme in Lille map}
	\label{fig4}
\end{figure*}

From Fig. ~\ref{fig4a} and Fig. ~\ref{fig4c}, we can observe that the performance between the two maps are different.
For example, the performance of the proposed scheme has deteriorated in the $Lille$ map.
This is due to the possibility of a more dynamic movement for the adversary, which can select sensor nodes from a wider range of choices compared to the line topology of the $Grenoble$ map.
This forces the adversary to only travel through the line, making the pattern more predictable and monotonous.
Even though the performance of the proposed scheme has deteriorated in the $Lille$ map, it still outperforms the other methods, where in the best case it can last about $4$ rounds more than the random method in reaching $20\%$ attack chance.
On the other hand, Fig. ~\ref{fig4b} shows a different performance trend compared to Fig. ~\ref{fig3b}, as depending on the starting location of the attackers, the movement can become redundant due to their circular search procedure.

Finally, we observe the $d(f_k)$ value, which represents the average distance between the fragments. 
Fig. ~\ref{fig5a} shows the average distance values from both $Grenoble$ and $Lille$ sites. 
The values of distances are normalized to the $Near-First$ method.
From the figure, we can observe that the proposed scheme maintains the highest value of average distance between fragments, which directly related to its security performance.
However, this will also increase the transmission overhead that will be required to transfer the fragments in a multi-hop manner.
The distance value is slightly higher for the $Lille$ map, as it can provide a more balanced form of clustering within the grid and fragments can be dispersed with more distance.

\subsection{Overhead evaluation of proposed scheme}

\begin{figure*}[!h]
	\begin{center}
		\subfigure[Average distance between fragments]{
			\label{fig5a}
			\includegraphics[width=5.8cm]{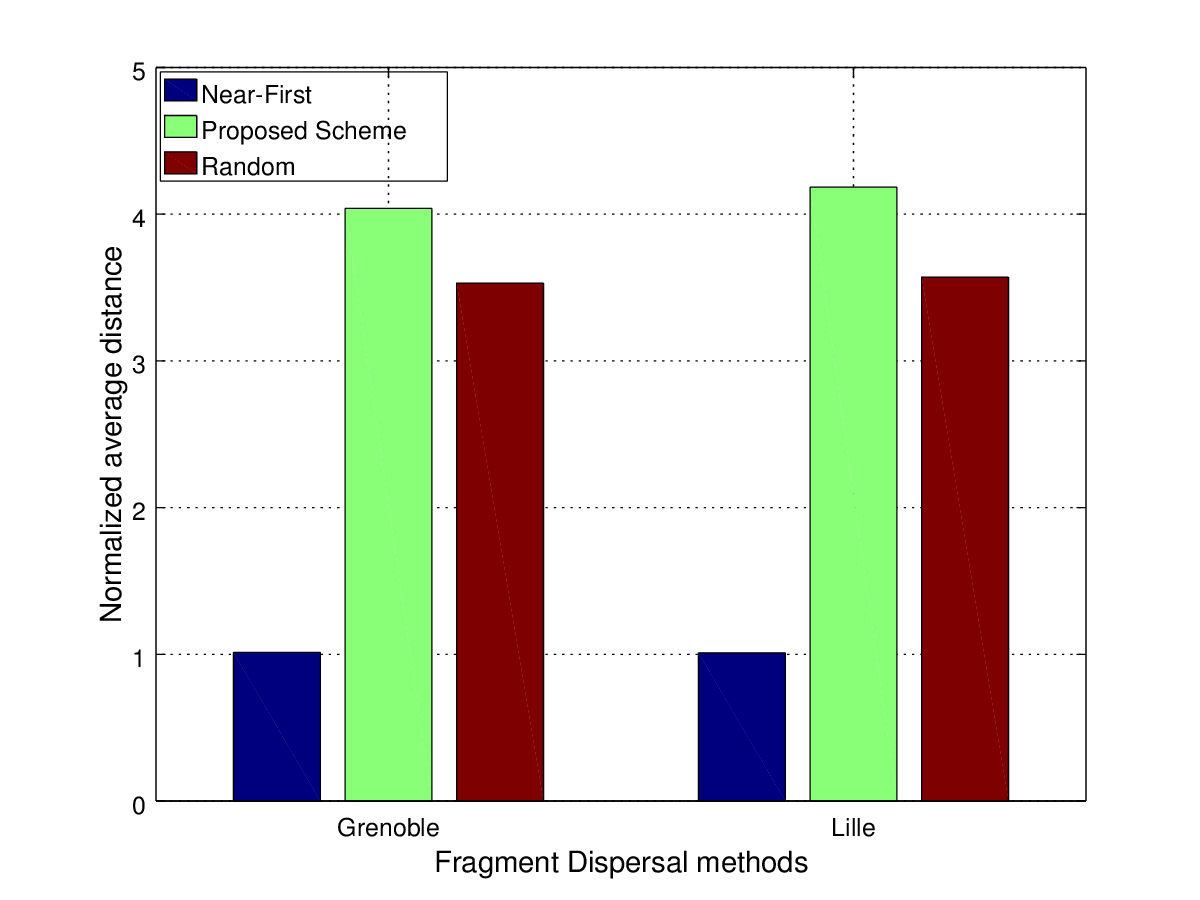}}
		\subfigure[Computational complexity of routing protocols]{
			\label{fig5b}		
			\includegraphics[width=5.8cm]{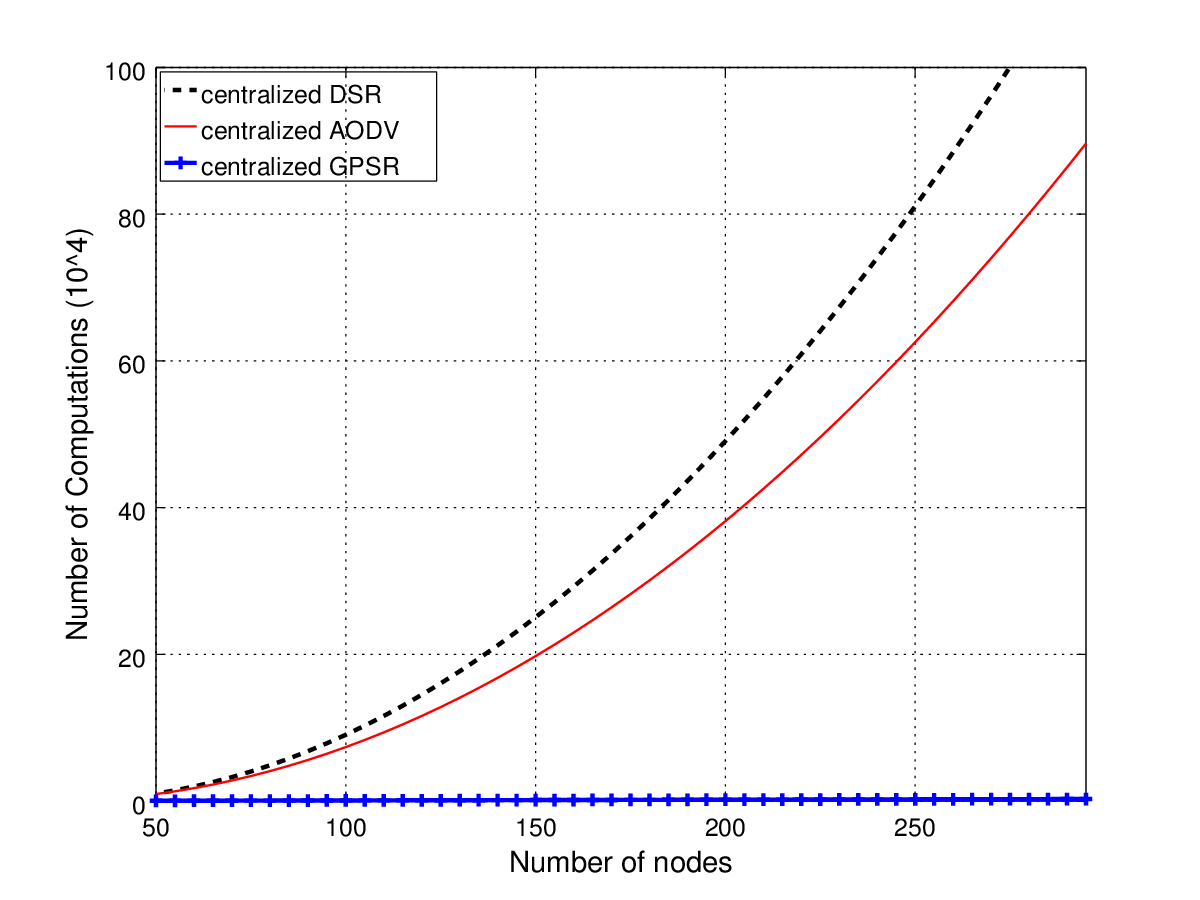}}	
		\subfigure[Communication complexity of routing protocols]{
			\label{fig5c}		
			\includegraphics[width=5.8cm]{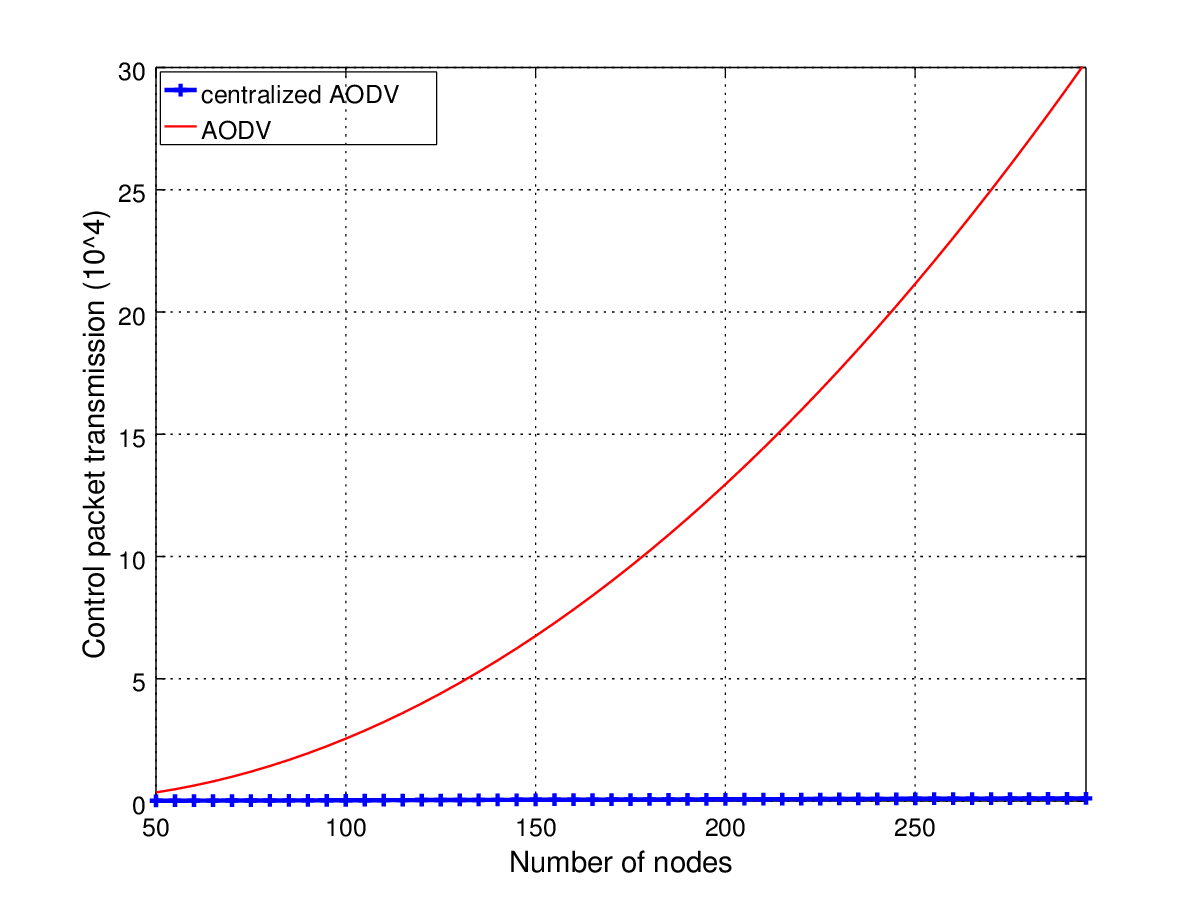}}	
		
	\end{center}
	\caption{Overhead evaluation of proposed scheme}
	\label{fig5}
\end{figure*}

From the security analysis, we have observed that our method of selecting multi-hop sensor nodes for fragment dispersal can provide higher security than other methods. 
However, multi-hop transfer of fragments require a usage of routing protocol inside the network, which is bound to consume energy.
As the fragmentation process itself increases the energy consumption, it is important to keep the overhead of the routing protocol as low as possible.
Here we make a mathematical analysis using MATLAB and compare the computational complexity of the routing protocols within the sink node and communication complexity for the sensor nodes.
Note that our objective is not to make a comparison of the overhead between the proposed scheme and $Near-First$ or $Random$ method, as it is intuitive that multi-hop communication will induce more overhead.
Instead, we emphasize that our main goal here is to reduce the overhead that is induced from the routing protocol.

Fig. ~\ref{fig5b} shows the computational complexity of the centralized routing protocols proposed in this work. 
The complexity is computed by counting the number of instructions executed by the itinerant sink.
We assume that the itinerant sink makes a calculation of the routes of all nodes every time it finishes a sink trip.
In the figure, we can observe that GPSR induces the lowest overhead among the protocols.
This is because GPSR does not induce any control overhead, using only the geographical information acquired from the clustering phase of the proposed scheme.
Therefore, we can conclude that although all the routing protocols can be used, GPSR could be favorable for specific applications where there are limitations in the sink, or if there are numerous amount of sensor nodes in the network.
Note that if traditional routing versions are used instead, they induce very little to no overhead at the mobile sink's side because the calculations are all done by sensor nodes in a distributed manner.
The load induced at the sink side is higher for the proposed centralized routing protocols.
However, we argue that the trade-off of increasing the computational load of sink to decrease the communication of sensor node is beneficial in UWSN, because in practice it is difficult to provide high-capacity energy sources to sensor nodes while mobile sink is easily retrievable and rechargeable.
Fig. ~\ref{fig5c} shows the communication overhead of control messages for the routing protocols.  
Here we assume that one sensor node creates one data per one itinerant sink trip.
Also, the route table timeout of AODV and DSR is shorter than one sink trip, so route requests must be made by each node per sink trip.
Here we only show the comparison of AODV and its centralized counterpart. 
This is because DSR has similar broadcast/unicast properties as AODV, while GPSR does not induce any control overhead.
In our case, we have utilized the itinerant sink to approximate their locations, as explained in the previous section.
In the figure, we can observe that the control overhead from route request and route reply has dramatically decreased.
In terms of reducing control overhead, it could be argued that only GPSR need to be used, because GPSR does not need any route request exchanges while AODV and DSR is heavy in terms of overhead.
However, we note that the routing paths created by GPSR is not shortest-hop, due to its greedy routing approach.
Therefore, depending on specific applications, AODV or DSR can also be used, in which case our method can clearly provide better utilization of these protocols.
As a conclusion, we sacrifice the resources of the itinerant sink, which does the jobs of the sensor nodes in a centralized manner.
Through this, we gain significant reduction in control overhead, which results in less energy consumption from packet transmission per node.

\section{Discussions}

Here in this section, we discuss some limitations of our work, and how we will try to tackle them in our future research.

\subsection{On the tradeoff between energy and data survivability}

As explained in our analysis, the distance between sensor nodes and the data originator causes additional transmission overhead. 
Therefore, there clearly exists a tradeoff between level of security (survivability of the data) and energy consumption. 
Although we have addressed the issues of reducing overhead in the routing protocol itself, we do not address this tradeoff in this work. 
This problem will be considered in our future work, as we believe the level of security needed in the network can be dynamically controlled by predicting and analyzing the current and future danger level of the network.

\subsection{On the concentration of fragments on edge sensor nodes}

In our current work, we utilize a random method to select a sensor within each area of the network. 
This was used because the method of choosing the most distant sensor nodes is not feasible as it may concentrate usage of energy to only the edge nodes, causing various aforementioned problems. 
We believe that within an area, there can be more efficient methods on selecting the most ideal sensor node not only based on $d(f_k)$, but also on other important factors such as residual energy, number of fragments already in storage, link quality, etc. 
How to intelligently utilize these parameters will be also a focus of our future work.

\subsection{On the usage of our attacker and sink model}

In our work, we model the mobility of the attacker and sink according to a realistic design and a specific environment case; for example a drone physically accessing the sensor node to destroy/exploit data. 
Note that this can be considered as a general scenario of UWSN, and we believe that our work is a general modeling and observation of such scenario.
However, depending on the development of attacking methods and applications in IoT environment, other numerous models of attacks and movements will be bound to occur in the future.
Even in such complex cases, we believe that we can utilize our current UWSN model as a benchmark, and enhance it for more complex scenarios and environments, to analyze further threats that may affect the reliability of UWSNs, 

\subsection{Enhancement to the routing protocols}

In our work, we have modified existing distributed ad hoc routing protocols to operate under centralized circumstances.
However, instead of just converting their computational properties, we believe that each routing protocol itself can be enhanced for UWSN environments.
For example, when selecting destinations for fragment dispersion, destinations can be selected via clusters, instead of selecting individual sensor nodes. 
This allows routing tables of AODV to operate in a more efficient way, as the entries in the routing table will never be higher than the number of fragments.
For the selection of sensor nodes inside a cluster to store the fragment, anycasting methods can be used to randomly select a node and forward the data inside the cluster.
%

GPSR also has a problem in its greedy forwarding process; if the node that needs to forward the data is the physically closest node to the destination, then it must return the data packet to the previous node.
Even though routing loop can be prevented through the right-hand rule, this still induces redundant data transmission.
However, in UWSN, as the itinerant sink has the global map of the network, the right-hand rule does not need to be used to avoid holes in the network.
This means that the holes in the network can be detected, which will allow the sink to pre-calculate the best routes to the destination node while avoiding network holes.
Therefore, GPSR can also become more efficient in terms of computational complexity within the itinerant sink's decision making process.

\section{Conclusion}

Unattended wireless sensor networks is an interesting and trendy area of research, which we believe can be used in many future applications and services, especially following the current trend of IoT. 
For its security, data fragmentation can be used, but how to place the fragments in a multi-hop manner is an important topic that needs to be resolved.
To solve this problem, we have designed an analytical model of a basic UWSN and presented some observations.
Furthermore, we argue that existing routing protocols cannot be used for UWSN, as they induce high level of communication overhead which is undesirable for energy-limited sensors.
Therefore, we modify the routing protocols to behave in a centralized manner, saving the energy cost of UWSN.

Through our analytical and experimental results, we show that our proposed scheme can provide more than $100\%$ performance increase in preventing attacks from adversaries.
Furthermore, compared to the traditional routing protocols, our proposed centralized routing approach can save energy consumed from route control overhead.
We also propose a method of approximating location of sensors using the itinerant sink, allowing geographical routing to be used.
We believe that the results we have acquired is promising, but we also believe that it can be further improved.
In the future, we will make more complex models of the sink/adversary to experiment with more practical environments.
We will also tune the routing protocols and enhance them to make them more suitable for data fragment dispersal.

\bibliographystyle{ACM-Reference-Format}
\bibliography{8-bibliography}

\end{document}